\documentclass[aps, prl, reprint, doublecolumn, nofootinbib, superscriptaddress]{revtex4-2}

\usepackage{amsmath}
\usepackage{amssymb}
\usepackage{amsfonts}
\usepackage{amsthm}
\usepackage{bm}
\usepackage{mathrsfs}
\usepackage{booktabs, array}
\usepackage{soul}
\usepackage[T1]{fontenc}
\usepackage[caption=false]{subfig}
\usepackage{graphicx} % Required for inserting images
\usepackage{capt-of}
\usepackage{multirow}
\usepackage{makecell}
\usepackage[dvipsnames]{xcolor}
\usepackage{tikz}
\usepackage[normalem]{ulem} % normalem prevents \emph from being underlined
\usetikzlibrary{positioning,fit,backgrounds,arrows.meta,calc}

\newcommand{\DD}{\boldsymbol{\Delta}}

\newcommand{\Gm}{\Gamma}
\newcommand{\Gb}{\bar\Gamma}
\newcommand{\Dop}[1]{D_{[#1]}}
\newcommand{\Nop}[1]{\DD_{[#1]}}
\newcommand{\dop}[1]{\delta_{[#1]}}
\newcommand{\dbop}[1]{\bar\delta_{[#1]}}
\newcommand{\lmax}{\ell_{\max}}

\definecolor{linkcolor}{rgb}{0.0,0.3,0.5}
\usepackage[
    hypertexnames=false,
    unicode,
    colorlinks=true,
    linkcolor=linkcolor,
    citecolor=linkcolor,
    filecolor=linkcolor,
    urlcolor=linkcolor,
    pdfusetitle
]{hyperref}
\usepackage{url}
\hypersetup{breaklinks=true}

\usepackage{orcidlink}
\newcommand{\dd}{\mathrm d}
\newcommand{\Real}{\operatorname{Re}}
\newcommand{\Pot}{\mathcal P}
\newcommand{\Src}{\mathcal S}

\newcommand{\UIUC}{Illinois  Center  for  Advanced  Studies  of  the  Universe 
\& Department of Physics, 
\\ University of Illinois Urbana-Champaign, Urbana, Illinois 61801, USA}

\allowdisplaybreaks
\begin{document}
    %TC:ignore
    \title{Analytic Metric for Rotating Black Holes in Higher-Derivative Gravity}
    \author{Jierui Hu}
    \affiliation{\UIUC}
    
    \author{Dongjun Li}
    \thanks{Contact author: \href{dongjun@illinois.edu}{dongjun@illinois.edu}}
    \affiliation{\UIUC}
    
    \author{Nicol\'{a}s Yunes}
    \affiliation{\UIUC}
    \begin{abstract}

    We turn metric reconstruction into an analytic solver for rotating black holes in modified gravity, bypassing the direct solution of coupled field equations. As an example, in parity-even cubic gravity, we construct the first analytic spinning black hole metric at linear order in the coupling, without spin or weak-field expansions. We stringently validate the field equations and reproduce independent analytic predictions for horizon properties. This establishes a new route to construct spinning black holes beyond Einstein gravity and lays the foundation for strong-field tests.
     
    \end{abstract}
    %TC:endignore
    \maketitle

%%%%%%%%%%%%%%%%%%%%%%%%%%%%%%%%%%%%%%%%%%%%%%%%%%%%%%%%%%%%%%
%%%%%%%%%%%%%%%%%%%%%%%%%%%%%%%%%%%%%%%%%%%%%%%%%%%%%%%%%%%%%%

\begin{figure*}[t]
    \centering
    \includegraphics[width=\textwidth]{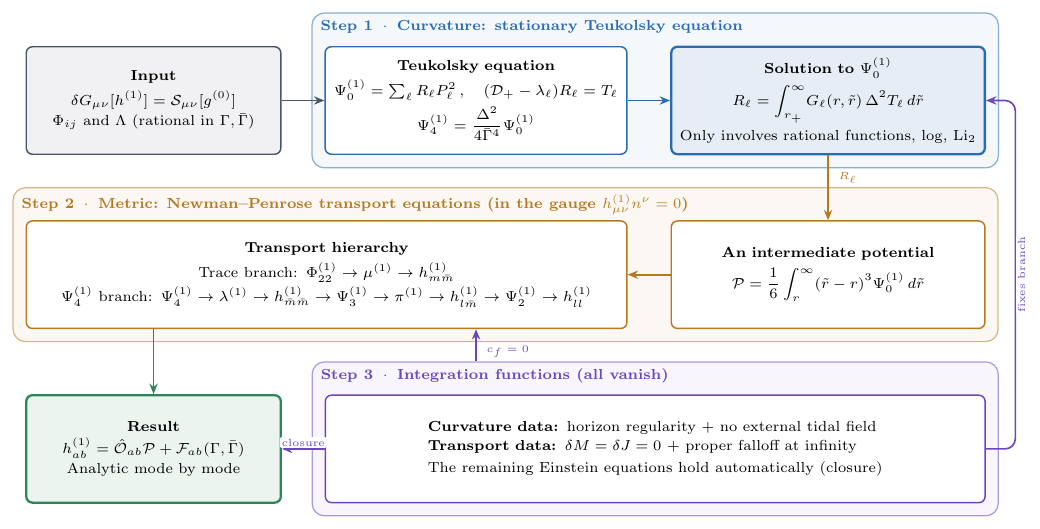}
    \caption{A flowchart for our analytic construction of the rotating black-hole metric in parity-even cubic gravity. The stationary Teukolsky equation, with mode source $T_\ell$, determines $\Psi_4^{(1)}$ as the input to the transport hierarchy of Ref.~\cite{Li:2026rkf}. The reconstructed metric correction $h_{ab}^{(1)}$ combines the curvature-driven contribution $\hat{\mathcal{O}}_{ab}\mathcal{P}$, evaluated mode by mode, with $\mathcal{F}_{ab}$, rational in $\Gamma$ and $\bar{\Gamma}$. Horizon regularity, asymptotic conditions, and fixed mass and angular momentum fix the integration functions.}
    \label{fig:flowchart}
\end{figure*}

%%%%%%%%%%%%%%%%%%%%%%%%%%%%%%%%%%%%%%%%%%%%%%%%%%%%%%%%%%%%%%
%%%%%%%%%%%%%%%%%%%%%%%%%%%%%%%%%%%%%%%%%%%%%%%%%%%%%%%%%%%%%%
\noindent \textit{Introduction---}Gravitational-wave observations test Einstein's general relativity in the strong-field, dynamical regime \cite{LIGOScientific:2025slb}. Despite its success in Solar-System \cite{Will:1993hxu} and gravitational-wave tests \cite{LIGOScientific:2021sio}, general relativity is often viewed as the low-energy limit of a more fundamental theory, such as string theory \cite{Schwarz:1982jn, Damour:1994zq, Banks:1996vh, Seiberg:1999vs, Aharony:1999ti, Mukhi:2011zz} or loop quantum gravity \cite{DeWitt:1967yk, DeWitt:1967ub, DeWitt:1967uc, DeWitt:1975ys, Birrell:1982ix, Ashtekar:1987gu, Rovelli:1989za, Rovelli:1997yv, Ashtekar:1997yu, Ashtekar:2004eh}, with possible departures organized as higher-derivative corrections in an effective field theory \cite{Donoghue:2012zc}. Under the no-hair theorems, stationary, asymptotically flat vacuum black holes in four-dimensional general relativity belong to the Kerr family \cite{Carter:1971zc, Robinson:1975bv}, whereas higher-derivative extensions can admit non-Kerr geometries \cite{Yunes:2024lzm, Yunes:2009hc, Yunes:2011we, Yagi:2012ya, Kleihaus:2011tg, Cano:2019ore, Fernandes:2025vxg, Lam:2025elw, Lam:2025fzi}. Accurate solutions for these geometries are essential for connecting strong-field observations to the underlying theories.

Early analytic black-hole solutions beyond Einstein gravity relied on nonrotating or slow-rotation approximations \cite{Yunes:2009hc,Yunes:2011we,Yagi:2012ya}, later extended to fourteenth order in spin and beyond~\cite{Cano:2019ore}. As observed in gravitational-wave detections \cite{LIGOScientific:2021djp} and predicted theoretically \cite{Buonanno:2007sv}, most of the merger remnants likely have spins $a\gtrsim0.7M$, where $M$ is the remnant mass, already motivating a move beyond slow rotation. Near-extremal black hole physics provides further motivation, as rapid rotation can amplify beyond-Einstein effects. The latter include large horizon tidal forces, potentially diverging in the extremal limit \cite{Horowitz:2023xyl, Horowitz:2024dch, Lam:2026grz}, and enhanced beyond-Einstein deviations of the ringdown spectrum near the transition between damped and zero-damping quasinormal modes \cite{Yang:2012he, Yang:2012pj, Cano:2025ejw}. Accurate solutions across the subextremal spin range, including near extremality, are, thus, important both observationally and theoretically.

Recent spectral and pseudospectral methods have made substantial progress toward this goal \cite{Kleihaus:2011tg, Fernandes:2025vxg, Lam:2025elw, Lam:2025fzi}. After solving the system at a sequence of fixed spins \cite{Fernandes:2025vxg, Lam:2025elw, Lam:2025fzi} and fitting their spin dependence, one can obtain convenient analytic representations of the numerically determined geometries \cite{Lam:2025elw, Lam:2025fzi}. Although highly useful for calculating observables, these representations retain numerical input, and resolving the near-extremal region can require increasing resolution and precision~\cite{Lam:2026grz}. A direct analytic construction, without a slow-rotation or a weak-field expansion, and without numerical fitting, would provide a complementary route to rapidly rotating black holes, while exposing structures that numerical representations may obscure.

We here turn the sourced metric reconstruction framework of Ref.~\cite{Li:2026rkf} into an analytic solver for rotating black holes beyond Einstein gravity. Treating the leading deviation from the Kerr metric as a stationary, axisymmetric perturbation, we reconstruct the corrected geometry from its perturbed curvature, rather than from directly solving the coupled metric field equations. Metric reconstruction \cite{Cohen_Kegeles_1975, Chrzanowski:1975wv, Kegeles:1979an, Lousto:2002em, Lousto:2005xu, Loutrel:2020wbw, Ripley:2020xby} and its sourced extensions \cite{Aksteiner:2016mol, Green:2019nam, Wardell:2024yoi} have been developed within black-hole perturbation theory, motivated in part by nonlinear ringdown \cite{Campanelli:1998jv, Ma:2024qcv, Khera:2024bjs} and higher-order self-force calculations \cite{Pound:2019lzj,Pound:2021qin,Wardell:2021fyy}. One of the main goals of this work is to repurpose and extend this machinery to construct the stationary black-hole geometry in modified theories of gravity.

Figure~\ref{fig:flowchart} summarizes our three-step framework. First, we solve the sourced, stationary Teukolsky equation mode by mode, obtaining the gauge-invariant Weyl perturbations $\Psi_0^{(1)}$ and $\Psi_4^{(1)}$. Second, we reconstruct the metric from $\Psi_4^{(1)}$ and the Ricci source using the Newman--Penrose transport equations~\cite{Li:2026rkf}. Their nested radial integrals reduce to a single potential, from which the metric follows by differentiation together with explicit source terms. Third, boundary conditions and the remaining Einstein equations fix the integration functions. The resulting metric is obtained analytically at first order in $\epsilon$, with the full spin- and strong-field dependence retained. At $\mathcal{O}(\epsilon)$, an exact identity expresses the field-equation residual of a truncated reconstruction entirely in terms of omitted source multipoles, allowing us to assess the accuracy of the method straightforwardly.

As a demonstration of this new method, we construct the first mode-by-mode analytic solution for rotating black holes in parity-even cubic gravity to first order in the coupling, valid in the strong-field regime and for generic subextremal spin. Numerical field-equation checks and agreement with independent horizon-thermodynamic predictions \cite{Reall:2019sah} validate the construction up to $a/M=0.9999$. Together, these results establish our approach as an analytic route to black-hole geometries beyond Einstein gravity and provide a foundation for strong-field tests of gravity.

%%%%%%%%%%%%%%%%%%%%%%%%%%%%%%%%%%%%%%%%%%%%%%%%%%%%%%%%%%%
%%%%%%%%%%%%%%%%%%%%%%%%%%%%%%%%%%%%%%%%%%%%%%%%%%%%%%%%%%%%%%
\vspace{0.2cm}
\noindent \textit{Parity-even cubic gravity and the stationary Teukolsky equation---}The reconstruction framework is best explained through a concrete example. For this, we choose parity-even cubic gravity, whose action is \cite{Cano:2019ore}:
\begin{equation} \label{eq:action}
    S=\frac{1}{16\pi}\int\dd^4x\sqrt{-g}
    \left[R+
    \epsilon R_{\mu\nu}{}^{\rho\sigma}R_{\rho\sigma}{}^{\alpha\beta}R_{\alpha\beta}{}^{\mu\nu}\right]\,,
\end{equation}
where $G=1=c$, $\epsilon=\lambda_{\rm ev}\ell_{\rm EFT}^4$, $\lambda_{\rm ev}$ is dimensionless, and $\ell_{\rm EFT}$ is the characteristic effective-field-theory length scale. The task is then to find a solution to the modified field equations associated with this action for a metric tensor $g_{\mu \nu}$ that represents a spinning black hole. We expand the stationary and axisymmetric metric as
\begin{equation}
    g_{\mu\nu}
    =g^{(0)}_{\mu\nu}+\epsilon h^{(1)}_{\mu\nu}
    +\mathcal O(\epsilon^2)\,,
    \label{eq:metric-decomposition}
\end{equation}
where $g^{(0)}_{\mu\nu}$ is the Kerr metric and $h^{(1)}_{\mu\nu}$ is the leading, even-parity cubic-gravity correction. Henceforth, all quantities with a superscript $(1)$ denote perturbations linear in $\epsilon$ from the Kerr background.

To begin the reconstruction, step 1 (see Fig.~\ref{fig:flowchart}) is to solve the sourced, stationary Teukolsky equation~\cite{Teukolsky:1973ha}. In the Kinnersley tetrad, this equation reads 
\begin{align} \label{eq:teuk}
    & \mathcal{H}_0\Psi_0^{(1)}
    =-43200 M^2a^2(1-x^2)\frac{\Gb}{\Gm^{11}}\,, \\
    & \mathcal{H}_0
    =\mathcal{D}_+
    +(1-x^2)\partial_x^2-2x\partial_x+2-\frac{4x^2}{1-x^2}\,,
\nonumber
\end{align}
where $x=\cos\theta$, $\Gm=r+iax$, $\Gb=r-iax$, $\Sigma=\Gm\Gb$, $\Delta=r^2-2Mr+a^2$, $r_\pm=M\pm\varsigma$, $\varsigma=\sqrt{M^2-a^2}$, and $\mathcal{D}_\pm\equiv\Delta^{\mp2}\partial_r\Delta^{1\pm2}\partial_r$. The source in Eq.~\eqref{eq:teuk} is driven by the effective stress-energy tensor associated with Eq.~\eqref{eq:action}, derived in the Supplemental Material. By using the discrete symmetry $(t,\phi)\mapsto(-t,-\phi)$ of Boyer--Lindquist coordinates, which interchanges the Kinnersley tetrad vectors $l^{\mu}\leftrightarrow n^{\mu}$ and $m^{\mu}\leftrightarrow\bar{m}^{\mu}$ up to rescaling, we find $\Psi_4^{(1)}$ is related to $\Psi_0^{(1)}$ via
\begin{equation} \label{eq:reflection}
    \Psi_4^{(1)}=\frac{\Delta^2}{4\Gb^4}\Psi_0^{(1)}\,,
\end{equation}
which supplies the curvature input for the reconstruction hierarchy in \cite{Li:2026rkf} in the traceful outgoing radiation gauge, $h^{(1)}_{\mu\nu}n^\nu=0$. The Kinnersley tetrad vectors $l^{\mu}$ and $n^{\mu}$ are the outgoing and ingoing principal null directions, respectively, with the remaining angular span $\{m^{\mu},\bar{m}^{\mu}\}$ fixed by the orthogonality conditions of the tetrad in the Newman--Penrose formalism \cite{Newman:1961qr, Chandrasekhar_1983}.

Since the spin-weight-$2$ spherical harmonics ${}_2Y_{\ell m}(\theta,\phi)$ solve the angular Teukolsky equation, we separate variables for our axisymmetric perturbation, $\Psi_0^{(1)}=\sum_{\ell\ge2}R_\ell(r)P_\ell^2(x)$, where $P_\ell^2(x)=(1-x^2)P_\ell''(x)\propto{}_2Y_{\ell0}$ are associated Legendre polynomials and primes denote derivatives with respect to the argument. With this decomposition, Eq.~\eqref{eq:teuk} becomes
\begin{subequations}
\begin{align}
    & (\mathcal D_+-\lambda_\ell)R_\ell=T_\ell\,,
    \label{eq:radial} \\
    & T_\ell=-43200 M^2a^2\nu_\ell
    \int_{-1}^{1}dx\,(1-x^2)^2P_\ell''(x)\frac{\Gb}{\Gm^{11}}\,,
    \label{eq:projection}
\end{align}
\end{subequations}
where $P_\ell$ are Legendre polynomials, $\lambda_\ell=(\ell-2)(\ell+3)$, $\nu_\ell=(2\ell+1)/(2K_\ell)$, and $K_\ell=(\ell-1)\ell(\ell+1)(\ell+2)$. 

The radial source $T_\ell$ comes from the projection of the right-hand side of Eq.~\eqref{eq:teuk} onto $P_{\ell}^2$. Since the integrand is a rational function of $x$, $T_\ell$ can be evaluated analytically at every $\ell$, yielding rational functions and logarithms. The homogeneous radial solutions to Eq.~\eqref{eq:radial} are $u_\ell=P_\ell''(y)$ and $v_\ell=Q_\ell''(y)$, where $y=(r-M)/\varsigma$ and $Q_\ell$ is the Legendre function of the second kind on its real exterior branch. The homogeneous solution $u_\ell$ is horizon-regular and behaves as $r^{\ell-2}$ at infinity, while $v_\ell$ decays as $r^{-\ell-3}$ but is horizon-singular. The horizon-regular solution without an external tidal field is then
\begin{equation} \label{eq:green}
    R_\ell
    =-\frac{1}{\varsigma^5K_\ell}\left(v_\ell\!\int_{r_+}^{r}
    u_\ell\Delta^2T_\ell\,d\tilde{r}
    +u_\ell\int_{r}^{\infty}v_\ell
    \Delta^2T_\ell\,d\tilde{r}\right)\,,
\end{equation}
where the weighted Wronskian is $\Delta^3(u_\ell\partial_r v_\ell-v_\ell\partial_r u_\ell)=-\varsigma^5K_\ell$. For every $\ell$, the rational and logarithmic structure of the integrands permits explicit evaluation, giving $R_\ell$ as a finite combination of rational functions, logarithms, and dilogarithms (see Supp.~Mat., Sec.~\ref{sm:integral-free}).

%%%%%%%%%%%%%%%%%%%%%%%%%%%%%%%%%%%%%%%%%%%%%%%%%%%%%%%%%%%%%%
%%%%%%%%%%%%%%%%%%%%%%%%%%%%%%%%%%%%%%%%%%%%%%%%%%%%%%%%%%%%%%
\vspace{0.2cm}
\noindent  \textit{Metric reconstruction via transport equations---}After obtaining $\Psi_4^{(1)}$, step 2 (see Fig.~\ref{fig:flowchart}) is to reconstruct the corresponding stationary metric perturbation. Following Ref.~\cite{Li:2026rkf}, we work in the traceful outgoing radiation gauge, $h^{(1)}_{\mu\nu}n^\nu=0$, where the nonzero metric components are $h_{ll}^{(1)}$, $h_{l\bar{m}}^{(1)}$, $h_{m\bar{m}}^{(1)}$, and $h_{\bar{m}\bar{m}}^{(1)}$. These components (and their complex conjugates) can be hierarchically solved from the sequence of first-order Newman--Penrose equations shown in the transport hierarchy box of Fig.~\ref{fig:flowchart} (see Supp.~Mat., Sec.~\ref{smtc:transport} for details). The quantities $\Psi_{2,3,4}$ are Weyl scalars, $\Phi_{22}$ is one of the Newman--Penrose Ricci scalars, and $\mu$, $\lambda$, $\pi$ are spin coefficients. 

The reconstruction route in Fig.~\ref{fig:flowchart} has two branches. The first branch determines the metric trace $h_{m\bar{m}}^{(1)}$ from $\Phi_{22}^{(1)}$, the latter of which can be computed from quantities on the Kerr background. The second branch determines the rest of the metric components directly from $\Psi_4^{(1)}$. Each arrow in the transport hierarchy box of Fig.~\ref{fig:flowchart} represents a first-order radial transport equation, with sources determined by preceding steps and the Ricci source. The principal part of each transport equation is $-(\Delta/2\Sigma)\partial_r$, so they can be solved analytically via an integrating factor $H_f$, which is a monomial in $\Delta$, $\Gm$ and $\Gb$, i.e.,
\begin{equation} \label{eq:transport}
    f=H_f\bigg[c_f(\theta)+2\int_r^\infty
    \frac{\Sigma \, S_f}{\Delta H_f}\,d\tilde r\bigg]\,,
\end{equation}
where $f$ is any of the nine quantities solved for along the two branches of Fig.~\ref{fig:flowchart}: $\mu^{(1)}$ and $h_{m\bar m}^{(1)}$ in the trace branch, and $\lambda^{(1)}$, $h_{\bar m\bar m}^{(1)}$, $\Psi_3^{(1)}$, $\pi^{(1)}$, $h_{l\bar m}^{(1)}$, $\Psi_2^{(1)}$, and $h_{ll}^{(1)}$ in the $\Psi_4^{(1)}$ branch. The quantity $S_f$ is the source in the transport equation, and $c_f$ is an integration function that we determine below. For example, the equations determining $h_{\bar m\bar m}^{(1)}$ are 
\begin{equation} \label{eq:shear}
    \partial_r\left(\frac{\Gb^4\lambda^{(1)}}{\Delta}\right)
    =\frac{\Sigma}{2}\Psi_0^{(1)}\,,\quad
    \partial_r\left(\frac{\Gb}{\Gm}\,h_{\bar{m}\bar{m}}^{(1)}\right)
    =-\frac{4\Gb^2}{\Delta}\lambda^{(1)}\,,
\end{equation}
where we have used the identity in Eq.~\eqref{eq:reflection} to replace $\Psi_4^{(1)}$ with $\Psi_0^{(1)}$, as we directly solve the latter from Eq.~\eqref{eq:green}. The solution to these equations for $h_{\bar{m}\bar{m}}^{(1)}$ can thus be cast in quadrature, through two nested integrals. The solution to the first-order transport equations that lead to the other projections of the metric perturbation can similarly be written in quadrature, in terms of nested integrals.

Cauchy's formula for repeated integration reduces the $\Psi_4^{(1)}$-driven nested integrals to radial moments $\mathcal I_n=\int_r^\infty(\tilde r-r)^n\Psi_0^{(1)}(\tilde r,\theta)\,d\tilde r/n!$. For the above example, the particular solution is $h_{\bar m\bar m}^{(1)}=-(2\Gm/\Gb^2)(2\mathcal I_2+\Gm\mathcal I_1)$. The $\Psi_3^{(1)}$ transport introduces $\mathcal I_3$ through a cubic radial weight $[\Gb+(\tilde r-r)]^3$, and explicit reduction of the remaining transport equations closes on $\mathcal I_0,\ldots,\mathcal I_3$ and their angular derivatives, without higher moments. Since $\partial_r\mathcal I_n=-\mathcal I_{n-1}$ for $n\ge1$, the Weyl-driven metric terms can therefore be obtained by differentiating a single potential,
\begin{equation}
    \Pot\equiv\mathcal I_3
    =\sum_{\ell\ge2}I_\ell(r)P_\ell^2(x)\,,\quad
    I_\ell(r)=\frac16\int_r^\infty(\tilde r-r)^3
    R_\ell(\tilde r)\,d\tilde r\,,
    \label{eq:potential}
\end{equation}
with $\Psi_0^{(1)}=\partial_r^4\Pot$, while the Ricci source contributes additional rational terms in $\Gm$ and $\Gb$. Although $\Pot$ resembles a Hertz potential \cite{Cohen_Kegeles_1975, Chrzanowski:1975wv, Kegeles:1979an, Wald:1978vm, Green:2019nam}, it is a calculational convenience rather than a prerequisite for our reconstruction approach. It obeys the stationary spin-$(-2)$ Teukolsky equation with a rational source whose fourth radial derivative gives the source in Eq.~\eqref{eq:teuk} (see Supp.~Mat., Sec.~\ref{sec:supp-source-moment}).

To obtain $\Pot$, all we need is $I_\ell$, and no additional integration of $R_\ell$ is required. Projecting the fourth radial antiderivative of the Teukolsky source onto $P_\ell^2$ gives $(\mathcal D_--k_\ell)I_\ell=\mathcal F_\ell$, where $k_\ell=(\ell-1)(\ell+2)$ and $\mathcal F_\ell=\int_r^\infty(\tilde r-r)^3T_\ell(\tilde r)\,d\tilde r/6$. Combining this with $\partial_r^4I_\ell=R_\ell$ and the identity $\Delta^2\partial_r^4=\mathcal D_-(\mathcal D_-+2)$, we obtain (see Supp.~Mat., Sec.~\ref{sec:supp-source-moment})
\begin{equation} \label{eq:adjoint}
    I_\ell
    =\frac{1}{K_\ell}
    \left(\Delta^2R_\ell-\mathcal{E}_\ell\right)\,,\quad
    \mathcal{E}_\ell
    =\big(\mathcal{D}_-+k_\ell+2\big)
    \mathcal{F}_\ell\,.
\end{equation}
Note that $\mathcal F_\ell$ contains only rational functions and logarithms, because this quantity is defined as an integral over $T_{\ell}$, which itself can be evaluated analytically to obtain rational functions and logarithms. Therefore, $I_\ell$ requires no special functions beyond those already present in $R_\ell$. 

Carrying out the above procedure, we finally obtain the stationary metric correction $h_{ab}^{(1)}$ in parity-even cubic gravity for a generic spin:
\begin{subequations} \label{eq:metric_final}
\begin{align}
    & h_{m\bar{m}}^{(1)}
    =-\frac{108 M^2}{35}
    \left[6\left(\Gm^{-6}+\Gb^{-6}\right)+S_{[1,5]}\right]\,,
    \label{eq:hmmc} \\
    & h_{\bar{m}\bar{m}}^{(1)}
    =-\frac{2\Gm}{\Gb^2}(\Gm\partial_r^2-2\partial_r)\Pot\,,
    \label{eq:hmcmc} \\
    & h_{l\bar{m}}^{(1)}
    =\frac{2\sqrt{2}\Gm}{\Gb^2\Delta}
    \left[\Sigma\mathscr{L}_2\partial_r
    -2r\mathscr{L}_2
    -j(\Gm-\Gb)\partial_r\right]\Pot\nonumber\\
    & \qquad+\frac{\sqrt2 j\Gm}{\Delta}G_R\,,
    \label{eq:hlmc} \\
    & h_{ll}^{(1)}
    =\frac{\Sigma^2}{\Delta^2}
    \left\{-8\Real\left[\frac1{\Gb^2}
    \Big(\mathscr{L}_1+\frac{2j}{\Gm}\Big)
    \mathscr{L}_2\Pot\right]+J_R\right\}\,,
    \label{eq:hll}
\end{align}  
\end{subequations}
where $j=ia\sin\theta$, $\mathscr{L}_s=\partial_\theta+s\cot\theta$, $S_{[m,n]}=\Gm^{-m}\Gb^{-n}+\Gm^{-n}\Gb^{-m}$, and
\begin{align}
    G_R
    =& \;\frac{M^2}{35}\left(\frac{368}{\Gm^6}
    +\frac{348}{\Gm^5\Gb}+\frac{40}{\Gm^4\Gb^2}
    +\frac{108}{\Gm\Gb^5}+\frac{648}{\Gb^6}\right)\,, \nonumber\\
    J_R
    =& \;\frac{16M^3}{35}(-8S_{[7,0]}-5S_{[6,1]}+27S_{[5,2]}) \nonumber\\
    & \;-\frac{16M^2a^2}{35}\left[92(1-x^2)S_{[7,1]}
    +(87-67x^2)S_{[6,2]}\right]\,.
 \label{eq:correctors}
\end{align}
The remaining components follow by complex conjugation or they vanish. Thus, the metric is obtained by differentiating the analytic potential $\Pot$ and adding explicit rational source terms. Contracting the metric with the dual Kinnersley tetrad, $h^{(1)}_{\mu\nu}=h^{(1)}_{ab}e^a_\mu e^b_\nu$ with $e^a_\mu=(-n_\mu,-l_\mu,\bar m_\mu,m_\mu)$ for $a=(l,n,m,\bar m)$, gives the metric in Boyer--Lindquist coordinates $(t,r,\theta,\phi)$, listed in Supp.~Mat., Sec.~\ref{smtc:bl-metric}. Although we only solve a subset of the Newman--Penrose equations here, we show in Supp.~Mat., Sec.~\ref{smtc:closure} that our procedure guarantees \textit{all} field equations are satisfied at $\mathcal O(\epsilon)$, following a similar proof to that presented in Ref.~\cite{Li:2026rkf}.

%%%%%%%%%%%%%%%%%%%%%%%%%%%%%%%%%%%%%%%%%%%%%%%%%%%%%%%%%%%%%%
%%%%%%%%%%%%%%%%%%%%%%%%%%%%%%%%%%%%%%%%%%%%%%%%%%%%%%%%%%%%%%
\vspace{0.2cm}
\noindent \textit{Fixing homogeneous solutions and integration functions---}The last step in the framework (see Fig.~\ref{fig:flowchart}) is to determine the homogeneous solutions and integration functions that appear in the metric reconstruction. In our procedure, there are two sets of homogeneous solutions or integration functions. The first set arises because each radial Teukolsky mode $R_\ell$ in Eq.~\eqref{eq:radial} carries two homogeneous solutions. Horizon regularity and the absence of an external tidal field exclude homogeneous terms proportional to $v_\ell$ and $u_\ell$, respectively, including the constant $\ell=2$ mode of $u_\ell$. 

The second set of integration functions arises because each application of Eq.~\eqref{eq:transport} introduces a homogeneous term $c_f(\theta)H_f$. With $\Psi_0^{(1)}=\Psi_4^{(1)}=0$ and vanishing Newman--Penrose Ricci scalars, the reconstruction hierarchy gives
\begin{align} \label{eq:homog}
    & \mathfrak{h}_{m\bar{m}}^{(1)}
    =c_{m\bar m}\Gb\Gm^{-1}
    +2c_\mu\left(\Gm^{-1}+\Gb^{-1}\right)\,,\nonumber\\
    & \mathfrak{h}_{\bar{m}\bar{m}}^{(1)}
    =c_{\bar m\bar m}\Gm\Gb^{-1}+4c_\lambda\Gm\Gb^{-2}\,,\nonumber\\
    & \mathfrak{h}_{l\bar{m}}^{(1)}
    =\Gm\Delta^{-1}\left[c_{l\bar m}-2c_\pi\Gm^2-\frac{4c_{\Psi_3}(\Gm+\Gb)}{3\Gb^2}\right]\,,\nonumber\\
    & \mathfrak{h}_{ll}^{(1)}
    =\Sigma^2\Delta^{-2}
    \left[c_{ll}\Gb\Gm^{-1}
    -4c_{\Psi_2}\left(\Gm^{-1}+\Gb^{-1}\right)\right]\,,
\end{align}
here $\mathfrak{h}^{(1)}_{ab}$ denotes the homogeneous metric perturbation. The quantities $c_{m\bar m}$, $c_{\bar m\bar m}$, $c_{l\bar m}$, and $c_{ll}$ are all functions of $\theta$ that arise from the solution to the metric transport equations, while $c_\mu,c_\lambda,c_{\Psi_3},c_\pi,c_{\Psi_2}$ arise from the integrations that lead to $\mu^{(1)},\lambda^{(1)},\Psi_3^{(1)},\pi^{(1)}$, and $\Psi_2^{(1)}$, respectively. 

To fix these quantities, we study how they affect the physics that our solution represents. First, we fix the ADM charges such that $\delta M=\delta J_{\rm ADM}=0$. Second, we require the perturbation to be mass-centered and to fall off as $h^{(1)}_{\hat a\hat b}=\mathcal{O}(r^{-3})$ in an asymptotically inertial frame, as a change of mass or spin would enter at $\mathcal{O}(r^{-1})$ or $\mathcal{O}(r^{-2})$. Expanding Eq.~\eqref{eq:homog} in $1/r$ at large $r$, we find that each radial function associated with the integration functions in $\mathfrak{h}_{m\bar{m}}^{(1)}$ and $\mathfrak{h}_{\bar{m}\bar{m}}^{(1)}$ enters at $\mathcal{O}(r^0)$ and $\mathcal{O}(r^{-1})$, so $c_{m\bar m}=c_\mu=c_{\bar m\bar m}=c_\lambda=0$. The leading term of the homogeneous perturbation $\mathfrak{h}_{l\bar{m}}^{(1)}$ goes as $-2c_\pi r$, followed by $c_{l\bar m}/r$ and $-8c_{\Psi_3}/(3r^2)$, which should also vanish. The leading term of the homogeneous perturbation $\mathfrak{h}_{ll}^{(1)}$ contains $c_{ll}r^0$ and $-8c_{\Psi_2}/r$, so $c_{ll}=c_{\Psi_2}=0$. Thus, all integration functions vanish under the chosen charge and asymptotic conditions. With these boundary conditions, the remaining Einstein equations are also satisfied at $\mathcal{O}(\epsilon)$ (see Supp.~Mat., Sec.~\ref{smtc:closure}).

\begin{figure*}[t]
    \centering
    \includegraphics[width=0.95\linewidth]{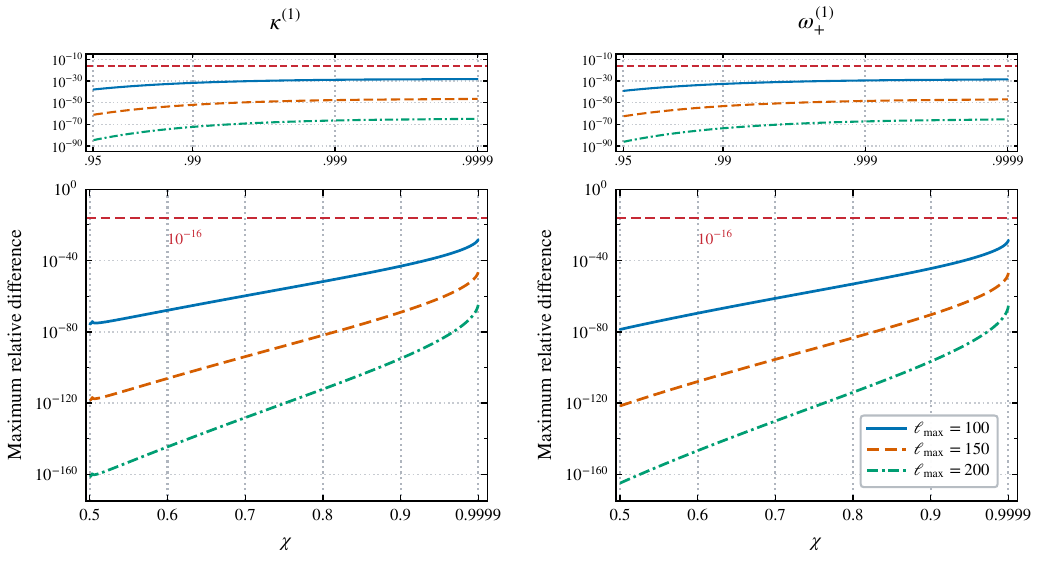}
    \caption{Maximum relative differences over five randomly chosen polar angles between metric-derived and analytic (see \cite{Reall:2019sah}) first-order corrections to the surface gravity, $\kappa^{(1)}$ (left), and horizon angular velocity, $\omega_+^{(1)}$ (right). The $1000$ spins span $0.5\leq\chi\leq0.9999$, and different colors denote the differences with $\ell_{\max}=100$ (blue), $150$ (orange), and $200$ (green), where the insets at the top resolve $\chi\geq 0.95$. Our horizon limits of $\kappa^{(1)}$ and $\omega_+^{(1)}$ are evaluated analytically. The red dashed lines mark the $10^{-16}$ double-precision reference.}
    \label{fig:cubic_horizon_quantities}
\end{figure*}

%%%%%%%%%%%%%%%%%%%%%%%%%%%%%%%%%%%%%%%%%%%%%%%%%%%%%%%%%%%%%%
%%%%%%%%%%%%%%%%%%%%%%%%%%%%%%%%%%%%%%%%%%%%%%%%%%%%%%%%%%%%%%
\vspace{0.2cm}
\noindent\textit{Validation of the analytic solution---}In practice, the mode representation of the potential $\mathcal{P}$ in Eq.~\eqref{eq:potential} must be truncated at some $\lmax$, and this does introduce some error in our solution. As we show below, however, the residual introduced due to truncation at $\lmax$ can be quantified \emph{analytically} and the solution converges exponentially with $\lmax$. Let us show this result by writing $\Pot=(1-x^2)F$ with $F=\sum_{\ell\ge2}I_\ell(r)P_\ell''(x)$, which then gives $\mathscr{Q}F=W$, where $\mathscr{Q}$ is a second-order partial differential operator and $W$ is rational in $\Gm$ and $\Gb$ (see Supp.~Mat., Sec.~\ref{sec:supp-validation}). Expanding $W=\sum_{\ell\ge2}\mathcal F_\ell(r)P_\ell''(x)$ gives $(\mathcal D_--k_\ell)I_\ell=\mathcal F_\ell$, so truncation replaces $W$ by its truncated expansion $W_{\lmax}$. The truncated metric satisfies the \emph{exact} identity
\begin{equation} \label{eq:identity}
    \delta G_{\mu\nu}[h_{\lmax}^{(1)}]-\Src_{\mu\nu}
    =\mathcal{C}_{\mu\nu}[W_{\lmax}-W]\,,
\end{equation}
where $\mathcal{C}_{\mu\nu}$ is a second-order, real-linear differential operator with rational Kerr coefficients, acting only on the omitted tail of $W$ (see Supp.~Mat., Sec.~\ref{sec:supp-validation}). 
At the retained coupling order, the analytic residual therefore arises solely from these omitted source multipoles. The source $W$ has a pole at $x=ir/a$ when analytically continued to the complex plane. 
This pole implies that when written in the separation-of-variables form shown above, its radial part $\mathcal F_\ell(r)$ has an exponentially decaying envelope $\varrho(r)^{-\ell}$ with $\ell$ (Supp.~Mat. Sec.~\ref{sec: angular convergence}), up to algebraic factors in $\ell$, where $\varrho(r)=(r+\sqrt{r^2+a^2})/|a|$. 
One calls this ``exponential convergence'' because, for fixed radius and spin, $\varrho$ is a constant, and thus $\varrho(r)^{-\ell}=\exp(-\ell \ln \varrho)$, which decays exponentially with $\ell$.
We see then that the argument of $\mathcal{C}_{\mu\nu}$ decays exponentially with $\ell$, and thus, the residual decays in the same fashion.  The angular convergence is slowest at the horizon, where $\varrho_{+}=(r_{+}+\sqrt{2Mr_{+}})/|a|$ decreases monotonically from infinity in the nonrotating limit to $1+\sqrt2$ at extremality, but it remains exponential, even at the horizon and as $|a|\to M$.

We now use the horizon limit of our analytic background metric correction $h_{ab}^{(1)}$ to compute the first-order corrections to the surface gravity, $\kappa^{(1)}$, and horizon angular velocity, $\omega_+^{(1)}$.
We compare these results with the independent predictions of Reall and Santos~\cite{Reall:2019sah}, who obtained these corrections from black hole thermodynamics without requiring the complete background metric. Figure~\ref{fig:cubic_horizon_quantities} shows the maximum relative differences over five randomly chosen polar angles for $1000$ spins in $0.5\le\chi\le0.9999$, where $\chi=a/M$. At $\chi=0.9999$, the maximum across both observables decreases from $6.2\times10^{-29}$ at $\lmax=100$ to $1.5\times10^{-65}$ at $\lmax=200$. Increasing $\ell_{\max}$ from $100$ to $200$ also reduces the differences between any two of the five sampled angles from $5.28\times10^{-28}$ to $1.24\times10^{-64}$ for $\kappa^{(1)}$ and from $2.65\times10^{-28}$ to $4.14\times10^{-65}$ for $\omega_+^{(1)}$, which is consistent with the expected constancy of $\kappa^{(1)}$ and $\omega_+^{(1)}$.

\begin{figure*}[t]
    \centering
    \includegraphics[width=0.95\linewidth]{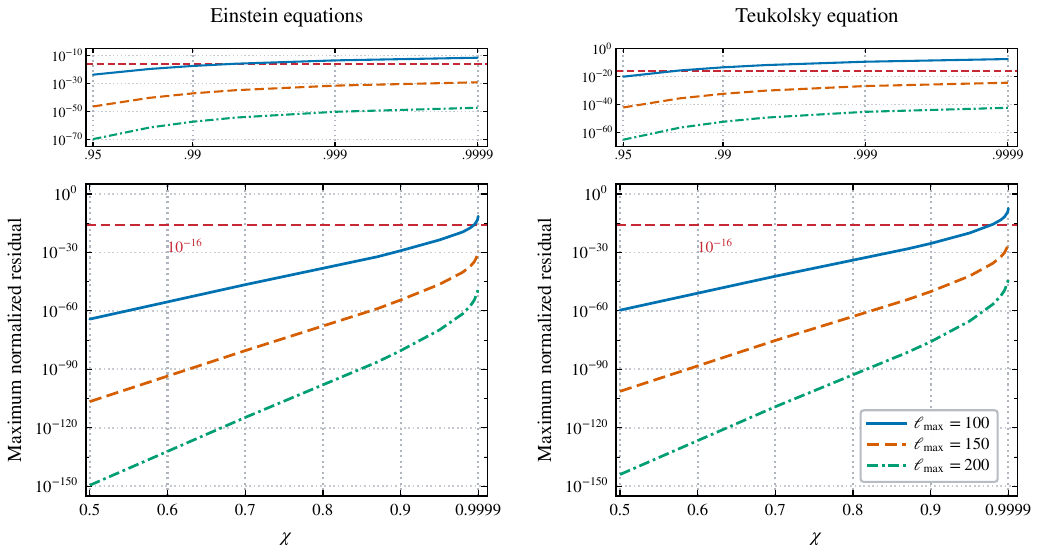}
    \caption{Maximum normalized residuals of the parity-even cubic-gravity metric for the Einstein equations (left) and the stationary Teukolsky equation of $\Psi_0^{(1)}$ (right). Each value is the maximum normalized residual over $200$ radii from $r=r_{+}+10^{-6}M$ to $r=100M$, with five randomly chosen polar angles $\theta$. Different colors denote the residuals at $\ell_{\max}=100$ (blue), $150$ (orange), and $200$ (green), where the insets at the top resolve $\chi\geq0.95$. The red dashed lines mark the $10^{-16}$ double-precision reference.}
    \label{fig:cubic_equation_checks}
\end{figure*}
To estimate how the truncation affects the accuracy to which our metric solves the modified field equations, we define the normalized field-equation residual
in the Kinnersley tetrad $e_a^\mu$ as
\begin{equation}\label{eq:norm}
    \mathcal{R}_{\lmax}
    =\frac{\max_{a,b}\left|e_a^{\mu} e_b^{\nu}
    \left(\delta G_{\mu\nu}[h_{\lmax}^{(1)}]-\Src_{\mu\nu}\right)\right|}
    {\max_{a,b}\left|e_a^\mu e_b^\nu\Src_{\mu\nu}\right|}\,,
\end{equation}
using the untruncated source in the denominator. We similarly normalize the stationary $\Psi_0^{(1)}$-Teukolsky residual by the source in Eq.~\eqref{eq:teuk}. At $1200$-digit working precision, we sample ten spins in $0.5\le\chi\le0.9999$, $200$ radii logarithmically spaced in $r-r_+$ from $r=r_++10^{-6}M$ to $r=100M$, with five fixed, randomly selected polar angles. 

Figure~\ref{fig:cubic_equation_checks} shows maxima over the sampled radii and angles. For $\lmax=100,150,200$, the field equation residuals are $5.5\times10^{-65}$, $2.7\times10^{-107}$, and $3.0\times10^{-150}$ at $\chi=0.5$ respectively, and $3.3\times10^{-12}$, $8.1\times10^{-30}$, and $4.3\times10^{-48}$ at $\chi=0.9999$. Adding $50$ multipoles gains approximately $42$--$43$ and $18$ decades in accuracy, respectively, consistent with our prediction $50\log_{10}\varrho_+\simeq44$ and $19$ up to algebraic finite-cutoff corrections. The Teukolsky-equation residual converges with similar features. The radial dependence of these residuals decreases strongly outward as $\varrho(r)$ increases, as we show in Fig.~\ref{fig:cubic_equation_checks_extended} of the End Matter. At $\chi=0.9999$ and $\lmax=200$, the field-equation and Teukolsky-equation residuals reach approximately $10^{-448}$ and $10^{-439}$, respectively, at $r=100M$. Near-horizon amplification in the chosen normalization makes the innermost samples the most stringent tests. Although $\lmax=50$ is insufficient there at the highest spins sampled (i.e., $\chi=0.9999$), both residuals fall below the $10^{-16}$ double-precision reference for $\lmax\ge150$ throughout the entire sampled grid.

%%%%%%%%%%%%%%%%%%%%%%%%%%%%%%%%%%%%%%%%%%%%%%%%%%%%%%%%%%%%%%
%%%%%%%%%%%%%%%%%%%%%%%%%%%%%%%%%%%%%%%%%%%%%%%%%%%%%%%%%%%%%%
\vspace{0.2cm}
\noindent\textit{Outlook---}We have turned metric reconstruction into an analytic solver for rotating black holes beyond Einstein gravity. In parity-even cubic gravity, this method yields a mode-by-mode purely analytic solution to first order in the coupling, which is valid for generic spin and does not employ a slow-rotation or weak-field expansion, nor numerical fitting. Truncating the analytic solution at a maximum mode number is our only error, but it can be characterized analytically, and the solution converges exponentially with mode number, even at near-extremal spins. Furthermore, this framework offers a new route to construct spinning black hole solutions analytically in other higher-derivative, effective field theories. We have already constructed analytic solutions in some of these theories using the framework built in this work, and a forthcoming catalog will present the full collection and detailed derivations \cite{Li:2026def}. These geometries provide the foundation for ringdown and extreme-mass-ratio-inspiral modeling, and for tests of gravity with black-hole images and shadows. Their analytic structure also enables systematic studies of near-extremal tidal effects, horizon thermodynamics, and the validity of the gravitational effective-field-theory expansion.

%TC:ignore
%%%%%%%%%%%%%%%%%%%%%%%%%%%%%%%%%%%%%%%%%%%%%%%%%%%%%%%%%%%%%%
%%%%%%%%%%%%%%%%%%%%%%%%%%%%%%%%%%%%%%%%%%%%%%%%%%%%%%%%%%%%%%
\vspace{0.2cm}
\noindent  \textit{Acknowledgment---}We thank Roman Berens and Kelvin Lam for insightful discussions. J.H., D.L., and N.Y. acknowledge support from the Simons Foundation (via Award No. 896696), the Simons Foundation International (via Grant No. SFI-MPS-BH-00012593-01), and the NSF (via Grant No. PHY-2512423). Codex (OpenAI) and Claude Code (Anthropic) were used to assist with code development and debugging, consistency checks, figure preparation, and English grammar corrections. All AI-assisted outputs have been independently verified by the authors.

\bibliographystyle{apsrev4-2}
\bibliography{references, supplement_reference}

% END MAIN TEXT
% BEGIN SUPPLEMENTAL MATERIAL
% The bibliography above is shared by the Letter and Supplemental Material.
\clearpage
\onecolumngrid

\begin{center}
{\large\bfseries End Matter}\\[4pt]
\end{center}

\noindent
\begin{minipage}{\textwidth}
    \centering
    \includegraphics[width=0.9\linewidth]{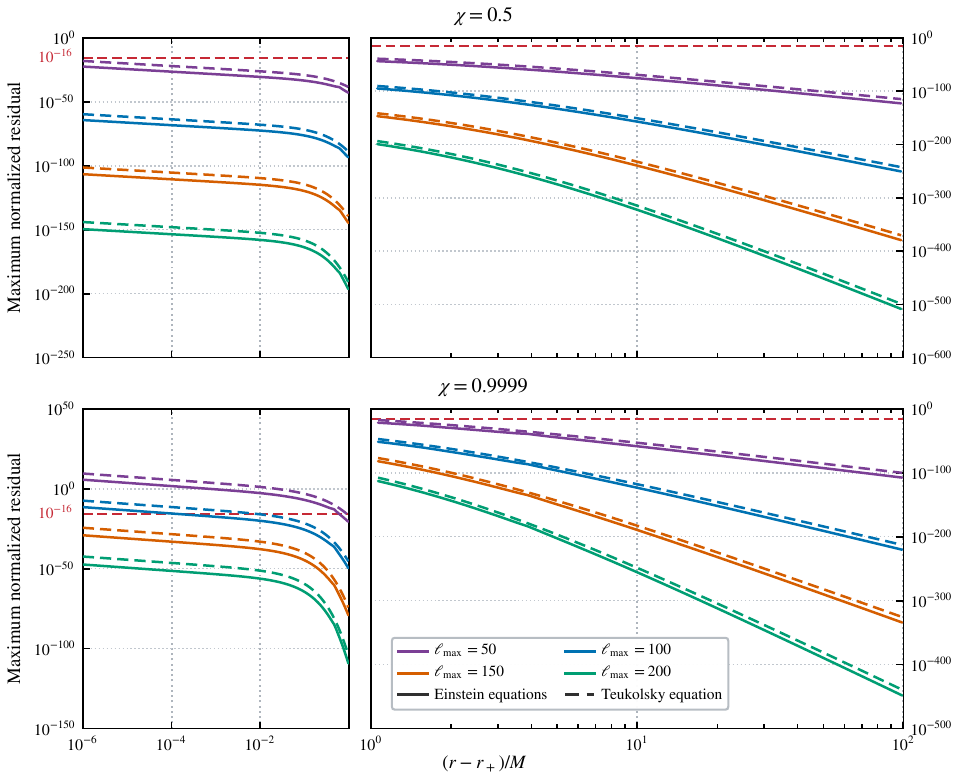}
    \captionof{figure}{Radial profiles of the normalized Einstein-equation (solid) and stationary $\Psi_0^{(1)}$-Teukolsky-equation (dashed) residuals for $\chi=0.5$ (upper row) and $\chi=0.9999$ (lower row). Each point is the maximum normalized residual over five randomly chosen polar angles $\theta$, and different line colors indicate $\ell_{\max}=50$ (purple), $100$ (blue), $150$ (orange), and $200$ (green), respectively. The logarithmic horizon-offset axis $(r-r_+)/M$ is split at unity, with an independent residual scale in each panel. The red dashed lines mark the $10^{-16}$ double-precision reference.}
    \label{fig:cubic_equation_checks_extended}
\end{minipage}

\setcounter{equation}{0}
\setcounter{section}{0}
\setcounter{subsection}{0}
\setcounter{table}{0}
\setcounter{figure}{0}
\renewcommand{\theequation}{S\arabic{equation}}
\renewcommand{\thesection}{S\Roman{section}}
\renewcommand{\thetable}{S\arabic{table}}
\renewcommand{\thefigure}{S\arabic{figure}}
% Distinct hyperlink destinations despite counter resets.
\providecommand{\theHequation}{}
\providecommand{\theHsection}{}
\providecommand{\theHsubsection}{}
\providecommand{\theHtable}{}
\providecommand{\theHfigure}{}
\renewcommand{\theHequation}{supp.\arabic{equation}}
\renewcommand{\theHsection}{supp.\arabic{section}}
\renewcommand{\theHsubsection}{supp.\arabic{section}.\arabic{subsection}}
\renewcommand{\theHtable}{supp.\arabic{table}}
\renewcommand{\theHfigure}{supp.\arabic{figure}}
\newpage
\begin{center}
{\large\bfseries Supplemental Material for}\\[4pt]
{\large\bfseries Analytic Metric for Rotating Black Holes in Higher-derivative Gravity} \\[6pt]
\end{center}
\bigskip

\setcounter{secnumdepth}{2}
This supplement collects the technical details behind the Letter.  Section~\ref{smtc:conventions} fixes conventions, defines the weighted NP derivatives, and lists the source. Section~\ref{sm:teukolsky} solves the stationary Teukolsky equation: the sources, the map from $\Psi_0^{(1)}$ to $\Psi_4^{(1)}$, the boundary-selected radial modes, and their closed form. Section~\ref{smtc:transport} integrates the transport hierarchy, introduces the reconstruction potential, and gives the explicit metric. Section~\ref{smtc:completion} fixes all integration functions, charges, and residual gauge, and proves closure of the remaining Einstein equations. Section~\ref{sec:supp-validation} derives the finite-cutoff Einstein identity.

% Preserve the same reference order as the Letter; every supplemental reference
% is also present in the Letter bibliography.
\nocite{Cano:2019ore,Fernandes:2025vxg,Lam:2025elw,Lam:2025fzi,Teukolsky:1973ha,Green:2019nam,Hollands:2024iqp,Li:2026rkf}

%%%%%%%%%%%%%%%%%%%%%%%%%%%%%%%%%%%%%%%%%%%%%%%%%%%%%%%%%%%%%%
%%%%%%%%%%%%%%%%%%%%%%%%%%%%%%%%%%%%%%%%%%%%%%%%%%%%%%%%%%%%%%
\section{Conventions, operators, and source}
\label{smtc:conventions}

We use the metric signature $(-+++)$ and the Riemann and Ricci tensor conventions
\begin{equation}
 R^\mu{}_{\nu\rho\sigma}=\partial_\rho\Gamma^\mu{}_{\sigma\nu}-\partial_\sigma\Gamma^\mu{}_{\rho\nu}
 +\Gamma^\mu{}_{\rho\lambda}\Gamma^\lambda{}_{\sigma\nu}-\Gamma^\mu{}_{\sigma\lambda}\Gamma^\lambda{}_{\rho\nu},
 \qquad R_{\nu\sigma}=R^\mu{}_{\nu\mu\sigma},
 \label{smtc:riemann}
\end{equation}
where $\Gamma^\mu{}_{\nu\rho}$ (with indices) are Christoffel symbols (not to be confused with the function $\Gm$, which does not carry indices). We use Greek indices for coordinate components, and Latin indices for tetrad (Newman--Penrose) components. The action and the perturbative expansion of the metric were given already in Eqs.~\eqref{eq:action} and \eqref{eq:metric-decomposition}, but we repeat them here for completeness and to introduce notation:
\begin{equation} \label{smtc:action}
    S=\frac{1}{16\pi}\int d^4x\sqrt{-g}\,(R+\epsilon I_3)\,,\qquad 
    I_3=R_{\mu\nu}{}^{\rho\sigma}R_{\rho\sigma}{}^{\alpha\beta}
    R_{\alpha\beta}{}^{\mu\nu}\,,\qquad 
    g_{\mu\nu}=g_{\mu\nu}^{(0)}
    +\epsilon h^{(1)}_{\mu\nu}+\mathcal O(\epsilon^2)\,.
\end{equation}
As in the Letter, the superscript $(1)$ denotes a first-order quantity, i.e., one multiplied by $\epsilon$. We keep it on the metric perturbation, the Weyl and NP Ricci scalars, the spin coefficients, and the transport sources, e.g., $h^{(1)}_{\mu\nu}$, $\Psi_{0,4}^{(1)}$, $\Phi_{ij}^{(1)}$, $\Lambda^{(1)}$, $\mu^{(1)}$, and $\Psi_2^{(1)}$, while background NP quantities, such as $\mu$ and $\Psi_2$, carry no superscript. Auxiliary functions built from first-order quantities, such as the radial modes $R_\ell$ and the potential $\mathcal P$, carry no superscript either. Evaluating the cubic terms on the Kerr metric gives the order-reduced equation
\begin{equation} \label{smtc:source}
    \delta{G}_{\mu\nu}[h^{(1)}]
    =\mathcal{S}_{\mu\nu}\,,\qquad
    \mathcal{S}_{\mu\nu}
    =3R_\mu{}^{\rho\sigma\lambda}R_{\sigma\lambda}{}^{\alpha\beta}R_{\alpha\beta\rho\nu}
    +\frac12g_{\mu\nu}I_3
    -6\nabla^\rho\nabla^\sigma(R_{\mu\rho\alpha\beta}R_{\nu\sigma}{}^{\alpha\beta})\,,
\end{equation}
where we canceled the common factor of $\epsilon$ on the left- and right-hand sides, and the superscript of $g_{\mu\nu}^{(0)}$ is ignored for simplicity without compromising comprehensibility. This expression is nothing but the pure parity-even case of Ref.~\cite{Cano:2019ore}. The Ricci tensor entering the NP equations is
\begin{equation} \label{smtc:trace-reversal}
    \mathcal{R}_{\mu\nu}
    =\mathcal{S}_{\mu\nu}-\frac{1}{2}g_{\mu\nu}\mathcal{S}\,,\qquad 
    \mathcal{S}
    =g^{\mu\nu}\mathcal{S}_{\mu\nu}\,.
\end{equation}

As in the main body of the Letter, we set
\begin{gather} 
    x=\cos\theta,\quad s=\sin\theta,\quad
    \Gm=r+iax,\quad \Gb=r-iax,\quad j=ias,\nonumber\\
    \Sigma=\Gm\Gb,\quad \Delta=r^2-2Mr+a^2,\quad
    r_\pm=M\pm\varsigma,\quad \varsigma=\sqrt{M^2-a^2}\,.\label{smtc:notation}
\end{gather}
Complex conjugation is taken at real $r,\theta,M,a$, so $\bar j=-j$, and we work at fixed $|a|<M$. In Boyer--Lindquist coordinates $(t,r,\theta,\phi)$, the Kinnersley tetrad is
\begin{equation} \label{smtc:tetrad}
    l^\mu=\frac{1}{\Delta}\left(r^2+a^2,\Delta,0,a\right)\,,\qquad
    n^\mu=\frac{1}{2\Sigma}\left(r^2+a^2,-\Delta,0,a\right)\,,\qquad
    m^\mu=\frac{1}{\sqrt2\Gm}\left(ias,0,1,i/s\right)\,,
\end{equation}
with $l_\mu n^{\mu}=-1$, $m_{\mu}\bar m^{\mu}=1$, and $\Psi_2^{(0)}=-M/\Gb^3$. On stationary, axisymmetric fields the directional derivatives are
\begin{equation}
 D=\partial_r,\qquad \DD=-\frac{\Delta}{2\Sigma}\partial_r,
 \qquad \delta=\frac{\partial_\theta}{\sqrt2\Gm},
 \qquad \bar\delta=\frac{\partial_\theta}{\sqrt2\Gb},
 \label{smtc:directions}
\end{equation}
and the nonvanishing background spin coefficients are
\begin{equation}
 \rho=-\frac1{\Gb},\quad \mu=-\frac{\Delta}{2\Gm\Gb^2},\quad
 \gamma=\mu+\frac{r-M}{2\Sigma},\quad
 \pi=\frac{j}{\sqrt2\Gb^2},\quad
 \tau=-\frac{j}{\sqrt2\Sigma},\quad
 \beta=\frac{\cot\theta}{2\sqrt2\Gm},\quad
 \alpha=\pi-\bar\beta,
 \label{smtc:spin-coefficients}
\end{equation}
while $\kappa=\sigma=\lambda=\nu=\varepsilon=0$. Following Eq.~(10) of Ref.~\cite{Li:2026rkf}, we define the weighted derivatives
\begin{equation} \label{smtc:weighted-operators}
\begin{aligned}
    \Dop{n_1,n_2,n_3,n_4}
    &=D+n_1\varepsilon+n_2\bar\varepsilon+n_3\rho+n_4\bar\rho\,,&\qquad
    \Nop{n_1,n_2,n_3,n_4}
    &=\DD+n_1\mu+n_2\bar\mu+n_3\gamma+n_4\bar\gamma\,, \\
    \dop{n_1,n_2,n_3,n_4}
    &=\delta+n_1\bar\alpha+n_2\beta+n_3\bar\pi+n_4\tau\,, &\qquad
    \dbop{n_1,n_2,n_3,n_4}
    &=\bar\delta+n_1\alpha+n_2\bar\beta+n_3\pi+n_4\bar\tau\,,
\end{aligned}
\end{equation}
for integers $n_1,\dots,n_4$. Since $\varepsilon=0$ in the Kinnersley tetrad, $\Dop{n_1,n_2,n_3,n_4}$ does not depend on $n_1$ and $n_2$. These four operators are used in all Newman--Penrose equations below.

The Ricci input from Eqs.~\eqref{smtc:source} and \eqref{smtc:trace-reversal} is
\begin{align} \label{smtc:np-source}
    & \Phi_{00}^{(1)}
    =324M^2(\Gm^{-8}+\Gb^{-8})\,,\qquad
    \Phi_{01}^{(1)}
    =54\sqrt2M^2j\left(3\Gm^{-1}\Gb^{-8}-\Gm^{-9}\right)\,,\qquad
    \Phi_{02}^{(1)}
    =-162M^2a^2s^2\left(\Gm^{-2}\Gb^{-8}+\Gm^{-10}\right)\,, \nonumber\\
    & \Phi_{11}^{(1)}
    =-27M^2(\Delta+a^2s^2)\Sigma^{-1}\left(\Gm^{-8}+\Gb^{-8}\right)\,,\qquad
    \Phi_{21}^{(1)}
    =-\frac{\Delta}{2\Gb^2}\Phi_{01}^{(1)}\,,\qquad
    \Phi_{22}^{(1)}
    =81M^2\Delta^2\left(\Gm^8+\Gb^8\right)\Sigma^{-10}\,, \nonumber\\
    & \Lambda^{(1)}
    =45M^2\left(\Gm^{-8}+\Gb^{-8}\right)
    -45M^3\left(\Gm^{-8}\Gb^{-1}+\Gb^{-8}\Gm^{-1}\right)
    -61M^3\left(\Gm^{-9}+\Gb^{-9}\right)\,,
\end{align}
with $\Phi_{ji}^{(1)}=\bar{\Phi}_{ij}^{(1)}$.

%%%%%%%%%%%%%%%%%%%%%%%%%%%%%%%%%%%%%%%%%%%%%%%%%%%%%%%%%%%%%%
%%%%%%%%%%%%%%%%%%%%%%%%%%%%%%%%%%%%%%%%%%%%%%%%%%%%%%%%%%%%%%
\section{Stationary Teukolsky equation}
\label{sm:teukolsky}

%%%%%%%%%%%%%%%%%%%%%%%%%%%%%%%%%%%%%%%%%%%%%%%%%%%%%%%%%%%%%%
%%%%%%%%%%%%%%%%%%%%%%%%%%%%%%%%%%%%%%%%%%%%%%%%%%%%%%%%%%%%%%
\subsection{Teukolsky equations and their sources}
\label{sm:teuk-sources}

For stationary, axisymmetric fields, the decoupled equations of
Ref.~\cite{Teukolsky:1973ha} for the extreme Weyl scalars reduce to
\begin{align} \label{sm:teuk-np}
    \left[\Dop{0,0,-4,-1}\Nop{1,0,-4,0}-\dop{-1,-3,1,-4}\dbop{-4,0,1,0}-3\Psi_2\right]\Psi_0^{(1)}
    & \equiv-\frac{1}{2\Sigma}\mathcal{H}_0\Psi_0^{(1)}\,, \nonumber\\
    \left[\Nop{4,1,3,-1}\Dop{0,0,-1,0}
    -\dbop{3,1,4,-1}\dop{0,4,0,-1}-3\Psi_2\right]\Psi_4^{(1)}
    & \equiv-\frac{1}{2\Sigma\Gb^4}
    \mathcal{H}_4\left(\Gb^4\Psi_4^{(1)}\right)\,,
\end{align}
where
\begin{equation} \label{sm:teuk-operators}
\begin{aligned}
    \mathcal H_0&=\mathcal D_+
    +(1-x^2)\partial_x^2-2x\partial_x+2-\frac{4x^2}{1-x^2}\,,\\
    \mathcal H_4&=\mathcal D_-
    +(1-x^2)\partial_x^2-2x\partial_x-2-\frac{4x^2}{1-x^2}\,,
\end{aligned}
\end{equation}
with $\mathcal D_\pm=\Delta^{\mp2}\partial_r\Delta^{1\pm2}\partial_r$ as in the main body of this Letter. In this separable form, $\mathcal{H}_4$ acts on $\Gb^4\Psi_4^{(1)}$ rather than on $\Psi_4^{(1)}$. The Bianchi identities supply the sources \cite{Li:2026rkf},
\begin{align} \label{sm:bianchi-sources}
    & S_0^{(1)}=\dop{-1,-3,1,-4}
    \left(\Dop{0,0,0,-2}\Phi_{01}-\dop{-2,-2,1,0}\Phi_{00}\right)
    +\Dop{0,0,-4,-1}\left(\dop{0,-2,2,0}\Phi_{01}
    -\Dop{0,0,0,-1}\Phi_{02}\right)\,, \nonumber\\
    & S_4^{(1)}=\Nop{4,1,3,-1}
    \left(\dbop{2,0,0,-2}\Phi_{21}-\Nop{0,1,2,-2}\Phi_{20}\right)
    +\dbop{3,1,4,-1}\left(\Nop{0,2,2,0}\Phi_{21}
    -\dbop{2,2,0,-1}\Phi_{22}\right)\,,
\end{align}
and the two equations become
\begin{equation}  \label{sm:teuk-pair}
    \mathcal{H}_0\Psi_0^{(1)}
    =\mathcal{T}_{2}\equiv-2\Sigma S_0^{(1)}\,,\qquad
    \mathcal H_4(\Gb^4\Psi_4^{(1)})
    =\mathcal{T}_{-2}\equiv-2\Sigma\Gb^4S_4^{(1)}\,.
\end{equation}
Inserting Eq.~\eqref{smtc:np-source} gives
\begin{equation} \label{sm:teuk-source}
    \mathcal{T}_{2}=-43200M^2a^2(1-x^2)\frac{\Gb}{\Gm^{11}}\,,\qquad
    \mathcal{T}_{-2}=\frac{\Delta^2}{4}\mathcal{T}_{2}\,.
\end{equation}
The source vanishes at $a=0$ and is invariant under $(a,x)\to(-a,-x)$, which leaves $\Gm$ and $\Gb$ unchanged.

%%%%%%%%%%%%%%%%%%%%%%%%%%%%%%%%%%%%%%%%%%%%%%%%%%%%%%%%%%%%%%
%%%%%%%%%%%%%%%%%%%%%%%%%%%%%%%%%%%%%%%%%%%%%%%%%%%%%%%%%%%%%%
\subsection{Relation between $\Psi_0^{(1)}$ and $\Psi_4^{(1)}$}
\label{smtc:reflection-derivation}

The horizon-regular, tide-free solutions of Eq.~\eqref{sm:teuk-pair} satisfy
\begin{equation}
    \Psi_4^{(1)}=\frac{\Delta^2}{4\Gb^4}\Psi_0^{(1)}\,,
    \label{smtc:reflection}
\end{equation}
with no complex conjugation. In Boyer--Lindquist coordinates, the Kerr metric contains $t$ and $\phi$ only through the $(t,t)$, $(t,\phi)$ and $(\phi,\phi)$ components. Hence,
\begin{equation}
    \iota:(t,r,\theta,\phi)\mapsto(-t,r,\theta,-\phi),\qquad
    \iota^*g^{(0)}=g^{(0)},\qquad \iota\circ\iota={\rm identity}\,,
    \label{smtc:iota}
\end{equation}
is an isometry. We use $\iota_*$ for the pushforward of vectors and $\iota^*$
for the pullback of covariant tensors, so at a spacetime point $P$,
$(\iota^*T)_P(X,Y,\dots)=T_{\iota(P)}(\iota_*X,\iota_*Y,\dots)$.
Then, it follows that $\iota_*$ reverses the $t$ and $\phi$ components of vectors and leaves $\Gm$, $\Gb$, $\Sigma$, and $\Delta$ unchanged, so the tetrad in Eq.~\eqref{smtc:tetrad} obeys
\begin{equation} \label{smtc:iota-tetrad}
    \iota_*l=-\frac{2\Sigma}{\Delta}\,n \,,\qquad
    \iota_*n=-\frac{\Delta}{2\Sigma}\,l \,,\qquad
    \iota_*m=\frac{\Gb}{\Gm}\,\bar m \,.
\end{equation}
Thus, $\iota$ makes the Geroch--Held--Penrose transformation $l\leftrightarrow n$, $m\leftrightarrow\bar m$ \cite{Geroch:1973am} and then applies a class-III tetrad rotation \cite{Chandrasekhar_1983}, i.e., a boost with parameter $A=-2\Sigma/\Delta$ and a spin by the phase $\Gb/\Gm$. The source in Eq.~\eqref{smtc:source} is built covariantly from $g_{\mu\nu}^{(0)}$ and is therefore $\iota$-invariant. By the uniqueness shown in Sec.~\ref{smtc:completion}, $\iota^*h=h$ up to gauge transformations. Since $\Psi_0^{(1)}$ and $\Psi_4^{(1)}$ vanish for the Kerr metric, their first-order perturbations are invariant under gauge transformations and tetrad rotations~\cite{Chandrasekhar_1983,Li:2022pcy}, so they are $\iota$-invariant functions of $(r,\theta)$. Given $\Psi_0^{(1)}=C^{(1)}_{\mu\nu\rho\sigma}l^\mu m^\nu l^\rho m^\sigma$ and $\Psi_4^{(1)}=C^{(1)}_{\mu\nu\rho\sigma}n^\mu\bar m^\nu n^\rho\bar m^\sigma$, where $C^{(1)}$ is the perturbed Weyl tensor, we can use $\iota^*C^{(1)}=C^{(1)}$, Eq.~\eqref{smtc:iota-tetrad}, and $\Psi_4^{(1)}(\iota(P))=\Psi_4^{(1)}(P)$ to get
\begin{equation} \label{smtc:reflection-proof}
    \Psi_0^{(1)}(P)
    =\left(\iota^*C^{(1)}\right)_P(l,m,l,m)
    =C^{(1)}_{\iota(P)}(\iota_*l,\iota_*m,\iota_*l,\iota_*m)
    =\frac{4\Sigma^2}{\Delta^2}\frac{\Gb^2}{\Gm^2}
    \,\Psi_4^{(1)}\big(\iota(P)\big)
    =\frac{4\Gb^4}{\Delta^2}\,\Psi_4^{(1)}(P)\,,
\end{equation}
which is Eq.~\eqref{smtc:reflection}. Equivalently, the radial operators obey $\mathcal H_4(\Delta^2f)=\Delta^2\mathcal{H}_0f$ for any $f(r,x)$, which together with $\mathcal T_{-2}=(\Delta^2/4)\mathcal{T}_{2}$ reproduces the same relation.

%%%%%%%%%%%%%%%%%%%%%%%%%%%%%%%%%%%%%%%%%%%%%%%%%%%%%%%%%%%%%%
%%%%%%%%%%%%%%%%%%%%%%%%%%%%%%%%%%%%%%%%%%%%%%%%%%%%%%%%%%%%%%
\subsection{Mode decomposition and the Green's function solution}
\label{sm:radial-branch}

With $\Psi_0^{(1)}=\sum_{\ell\geq2}R_\ell(r)P_\ell^2(x)$ and $P_\ell^2=(1-x^2)P_\ell''$,
the radial equation in Sturm--Liouville form is
\begin{equation}
 (\Delta^3R_\ell')'-\lambda_\ell\Delta^2R_\ell=\Delta^2T_\ell\,,
 \qquad
 T_\ell(r)=\nu_\ell\int_{-1}^{1}P_\ell^2(x)\,\mathcal T_2(r,x)\,dx\,,
 \label{smtc:radial-equation}
\end{equation}
where $\lambda_\ell=(\ell-2)(\ell+3)$, $\nu_\ell=(2\ell+1)/(2K_\ell)$,
$K_\ell=(\ell-1)\ell(\ell+1)(\ell+2)$, and $2K_\ell/(2\ell+1)$ is the norm of
$P_\ell^2$.

With $y=(r-M)/\varsigma$, the homogeneous solutions are $u_\ell=P_\ell''(y)$, which
is regular at $r_+$ and grows as $r^{\ell-2}$, and $v_\ell=Q_\ell''(y)$, with $Q_\ell$ being the real exterior branch of the Legendre functions of the second kind, which decays as $r^{-\ell-3}$ and is singular at the horizon $r_+$, where it has poles and a $\log(r-r_+)$ behavior. Their Wronskian is $\Delta^3(u_\ell v_\ell'-u_\ell'v_\ell)=-\varsigma^5K_\ell$, and variation of parameters
gives the particular solution
\begin{equation}
    R_\ell(r)=-\frac{1}{\varsigma^5K_\ell}\left[v_\ell(r)\int_{r_+}^{r}u_\ell\Delta^2T_\ell\,d\tilde r
    +u_\ell(r)\int_{r}^{\infty}v_\ell\Delta^2T_\ell\,d\tilde r\right]\,.
 \label{sm:green}
\end{equation}
 
We require the solution to be smooth at the future horizon and decay at spatial infinity, so there is no external tidal field. The particular solution in Eq.~\eqref{sm:green} satisfies both conditions: at large $r$ it falls off as $v_\ell\sim r^{-\ell-3}$, and at $r_+$ the factor $\int_{r_+}^r u_\ell\Delta^2T_\ell=\mathcal{O}((r-r_+)^3)$ removes the poles of $v_\ell$, while the remaining logarithms cancel between the two terms, so that $R_\ell$ is the analytic Frobenius branch (Sec.~\ref{sm:integral-free}). Every other solution differs from it by a homogeneous solution,
\begin{equation}
    R_\ell+\alpha_\ell\,u_\ell+\beta_\ell\,v_\ell\,,
    \label{smtc:curvature-uniqueness}
\end{equation}
where $\alpha_\ell$ and $\beta_\ell$ are constants. Smoothness at $r_+$ requires $\beta_\ell=0$, because $v_\ell$ is singular there, while $R_\ell$ and $u_\ell$ are regular. Decay at spatial infinity then requires $\alpha_\ell=0$ since $u_\ell$ grows as $r^{\ell-2}$. The solution is then uniquely given by Eq.~\eqref{sm:green}.

%%%%%%%%%%%%%%%%%%%%%%%%%%%%%%%%%%%%%%%%%%%%%%%%%%%%%%%%%%%%%%
%%%%%%%%%%%%%%%%%%%%%%%%%%%%%%%%%%%%%%%%%%%%%%%%%%%%%%%%%%%%%%
\subsection{Closed-form radial modes}
\label{sm:integral-free}

Both integrals in Eq.~\eqref{sm:green} can be carried out in closed form at every fixed $\ell$. The result is that each $R_\ell$ is a finite combination of rational functions, three logarithms, and four dilogarithms. We take $a>0$, so negative spins follow from $R_\ell(-a)=(-1)^\ell R_\ell(a)$. The construction has four steps: (i) evaluate the projected source, (ii) write the homogeneous solutions as polynomials plus one logarithm, (iii) compute the two antiderivatives required by Eq.~\eqref{sm:green} by finite algebraic operations, and
(iv) assemble the mode.

We use the following dimensionless coordinates and quantities for simplicity: 
\begin{equation} \label{sm:scaled-variables}
    z=\frac ra\,,\quad \hat M=\frac Ma\,, \quad
    b=\sqrt{\hat M^2-1}\,,\quad z_\pm=\hat M\pm b\,, \quad
    \hat\Delta=z^2-2\hat Mz+1=(z-z_+)(z-z_-)\,,\quad S=1+z^2\,,
\end{equation}
so that $\Delta=a^2\hat\Delta$, $z_+z_-=1$, and the horizon is at $z=z_+$. Primes on functions of $z$ denote $z$ derivatives. Writing
\begin{equation} \label{sm:mode-normalization}
    R_\ell(r)
    =\mathcal N_\ell\widehat R_\ell(z)\,,\qquad
    \mathcal N_\ell
    =-43200\frac{M^2}{a^{8}}\frac{2\ell+1}{2K_\ell}\,,
\end{equation}
the radial equation becomes
\begin{equation} \label{sm:unit-radial}
    \hat\Delta\widehat R_\ell''+6(z-\hat M)\widehat R_\ell'
    -\lambda_\ell\widehat R_\ell=J_\ell(z)\,,\qquad
    J_\ell(z)=\int_{-1}^{1}(1-x^2)^2P_\ell''(x)
    \frac{z-ix}{(z+ix)^{11}}\,dx\,.
\end{equation}
Integrate $J_\ell$ gives
\begin{equation} \label{sm:source-endpoint}
    J_\ell(z)
    =\frac{N_\ell(z)}{S^8}+B_\ell(z)L(z)\,,\quad
    L(z)=\log\frac{z+i}{z-i}\,,
\end{equation}
where $N_\ell$ and $B_\ell$ are polynomials. The denominator $S^{8}$ can be understood as follows. For $z=\pm i+\varepsilon$, the factor $z+ix$ becomes $\mathcal O(\varepsilon)$ at $x=\mp1$, where $(1-x^2)^2$ has a double zero, so the integrand behaves as $(1\pm x)^2/[\pm i(1\pm x)+\varepsilon]^{11}$. Changing the integration variable to $(1\pm x)/\varepsilon$ gives $J_\ell=\mathcal O(\varepsilon^{-8})$, so $N_\ell$ should be divided by $S^8$. Since the polynomial in the integrand of $J_\ell$ has degree $\ell+3$, $B_\ell=0$ for $2\leq\ell\leq6$. Angular orthogonality removes the first $\ell-2$ powers of $x$ in the large-$z$ expansion of the integrand, so $J_\ell=\mathcal O(z^{-\ell-8})$. The large-$z$ behavior of $J_\ell$ will be used to analyze the convergence of Eq.~\eqref{sm:first-antiderivative-algorithm}.

Now we examine the algebraic structure of the homogeneous solutions.
With $\xi=(z-\hat M)/b=(r-M)/\varsigma$, the growing solution $u_\ell=P_\ell''(\xi)$ is a polynomial. The decaying solution $v_\ell=Q_\ell''(\xi)$ follows from the standard form of the Legendre function of the second kind for $\xi>1$, $Q_\ell(\xi)=\tfrac12P_\ell(\xi)\log[(\xi+1)/(\xi-1)]-\omega_\ell(\xi)$, where $\omega_\ell$ is a polynomial. Since $(\xi+1)/(\xi-1)=(z-z_-)/(z-z_+)$, differentiating twice gives a polynomial times one logarithm plus a rational function. After rescaling by $b^{\ell-2}$, which makes all coefficients polynomial in $\hat M$ and $b^2$,
\begin{equation} \label{sm:homogeneous-scaled}
    \hat u\equiv b^{\ell-2}u_\ell=b^{\ell-2}P_\ell''(\xi),\qquad
    \hat v\equiv b^{\ell-2}v_\ell=\frac12\hat u\,H+\frac{b\,Y_\ell}{\hat\Delta^2}\,,\qquad
    H(z)=\log\frac{z-z_-}{z-z_+}\,,
\end{equation}
with the polynomials
\begin{equation} \label{sm:W-polynomial}
    Y_\ell(z)
    =b^{\ell+1}\left[-2(\xi^2-1)P_\ell'(\xi)+2\xi P_\ell(\xi)
    -(\xi^2-1)^2\omega_\ell''(\xi)\right]\,,\qquad
    \omega_\ell(\xi)
    =\sum_{k=1}^{\ell}\frac{P_{k-1}(\xi)P_{\ell-k}(\xi)}{k}\,.
\end{equation}
The Wronskian of $\hat{u}$ and $\hat{v}$ is $\mathcal W_\ell=\hat\Delta^3(\hat u\hat v'-\hat u'\hat v)=-K_\ell b^{2\ell+1}$.

Consider the integral
\begin{equation} \label{sm:first-antiderivative-algorithm}
    A_{\hat u}(z)\equiv\int_\infty^z \hat u(\tilde z)\hat\Delta^2(\tilde z)\,J_\ell(\tilde z)\,d\tilde z
    =a_{\hat u}(z)+f_{\hat u}(z)\,L(z)\,,
\end{equation}
where $f_{\hat u}(z)$ is a polynomial and $a_{\hat u}(z)$ is a polynomial divided by $S^7$, given that the rational part of $J_{\ell}(z)$ in Eq.~\eqref{sm:source-endpoint} is $N_\ell(z)/S^8$. Because the homogeneous solution $u_\ell$ grows as $z^{\ell-2}$, and $J_\ell=\mathcal O(z^{-\ell-8})$ in the large-$z$ limit, $A_{\hat u}(z)$ converges. The dimensionless form of the first integral in Eq.~\eqref{sm:green} can then be written as 
\begin{equation}
    \int_{z_+}^z \hat u(\tilde z)\hat\Delta^2(\tilde z)
    J_\ell(\tilde z)\,d\tilde z
    =A_{\hat u}(z)-A_+\,,
    \qquad A_+=A_{\hat u}(z_+)=a_{\hat u}(z_+)+f_{\hat u}(z_+)\log\frac{z_++i}{z_+-i}\,.
\end{equation}
The dimensionless form of the second integral in Eq.~\eqref{sm:green} is
\begin{equation} \label{sm:second-antiderivative-def}
    \int_{\infty}^z \hat v(\tilde z)\hat\Delta^2(\tilde z)
    J_\ell(\tilde z)\,d\tilde z
    =\frac12H A_{\hat u}+bE\,,\qquad
    E(z)\equiv\int_\infty^z
    \Big(Y_\ell J_\ell+\frac{A_{\hat u}}{\hat\Delta}\Big)dz'\,.
\end{equation}
To obtain Eq.~\eqref{sm:second-antiderivative-def}, we have used the relation $\hat v=\tfrac12\hat uH+bY_\ell/\hat\Delta^2$ from Eq.~\eqref{sm:homogeneous-scaled} and integrated the first term by parts with $H'=-2b/\hat\Delta$.

To evaluate $E$, we make an expansion of $f_{\hat u}$, i.e.,
\begin{equation}
    f_{\hat u}
    =\hat\Delta\left(f_{\rm q}+\frac{1}{2b}
    \left[\frac{f_{\hat u}(z_+)}{z-z_+}
    -\frac{f_{\hat u}(z_-)}{z-z_-}\right]\right)\,,
\end{equation}
where $f_{\rm q}$ is a polynomial. The integrand then splits into
\begin{equation} \label{eq:E integrand}
    Y_\ell J_\ell+\frac{A_{\hat u}}{\hat\Delta}
    =\underbrace{\frac{Y_\ell N_\ell}{S^8}
    +\frac{a_{\hat u}}{\hat\Delta}}_{\rm rational}
    +\underbrace{\left(Y_\ell B_\ell
    +f_{\rm q}\right)L}_{\text{polynomial}\,\times L}
    +\frac{1}{2b}\left[\frac{f_{\hat u}(z_+)}{z-z_+}-\frac{f_{\hat u}(z_-)}{z-z_-}\right]L\,,
\end{equation}
where the last term is the partial-fraction form of $f_{\rm r}L/\hat\Delta$, with $2b=z_+-z_-$. Since the denominators of the rational group above are $S^8$ and $S^7\hat{\Delta}$, the antiderivative of the rational group + polynomial$\times L$ group is
\begin{equation}
    r_E(z)+ l_E (z) L+h_E H+g_EG, \qquad G=\log(\hat\Delta/S),
\end{equation}
where $r_E(z)$ is a rational function and $l_E(z)$ is a polynomial. The quantities $h_E$ and $g_E$ are constants, since they come only from residues of simple poles. The last group in Eq.~\eqref{eq:E integrand} requires the dilogarithm pairs
\begin{equation} \label{sm:dilog-pair}
    \Phi_h(z)\equiv\int_\infty^z\frac{L(z')}{z'-h}\,dz'
    =\operatorname{Li}_2\!\left(\frac{-i-h}{z-h}\right)
    -\operatorname{Li}_2\!\left(\frac{i-h}{z-h}\right)\,,
    \qquad h=z_\pm\,.
\end{equation}
Altogether,
\begin{equation} \label{sm:second-antiderivative}
    E=r_E(z)+ l_E (z) L+h_E H+g_EG
    +\frac{f_{\rm r}(z_+)}{2b}\Phi_{z_+}
    -\frac{f_{\rm r}(z_-)}{2b}\Phi_{z_-}\,.
\end{equation}
Hence, Eq.~\eqref{sm:green} becomes
\begin{equation}
    \widehat R_\ell=\frac{\hat v\int_{z_+}^z \hat u(\tilde z)\hat\Delta^2(\tilde z)
    J_\ell(\tilde z)\,d\tilde z-\hat u\int_{\infty}^z \hat v(\tilde z)\hat\Delta^2(\tilde z)
    J_\ell(\tilde z)\,d\tilde z}{\mathcal W_\ell}
    =\frac{bY_\ell\hat\Delta^{-2}(A_{\hat u}-A_+)-\tfrac12\hat uH A_+-b\hat uE}{\mathcal W_\ell}\,.
 \label{sm:radial-closed}
\end{equation}
The closed form~\eqref{sm:radial-closed} is regular at the horizon. First, the integrand $\hat u\hat\Delta^2J_\ell$ vanishes as $(z-z_+)^2$, so $A_{\hat u}-A_+=\mathcal{O}((z-z_+)^3)$, which cancels the double pole of $\hat v$ in the first term. Second, the two remaining terms each contain $\log(z-z_+)$, and these cancel. Near $z_+$, $H=-\log(z-z_+)+\mathcal{O}(1)$, so $-\tfrac12\hat uHA_+\supset\tfrac12\hat uA_+\log(z-z_+)$. In the integrand of $E$, the term $Y_\ell J_\ell$ is regular at $z_+$, since $J_\ell$ has poles only at $z=\pm i$, whereas $A_{\hat u}/\hat\Delta$ has a simple pole with residue $A_{\hat u}(z_+)/(z_+-z_-)=A_+/(2b)$. Hence $E=(A_+/2b)\log(z-z_+)+\mathcal{O}(1)$, and $-b\hat uE\supset-\tfrac12\hat uA_+\log(z-z_+)$, which cancels the logarithm above. Thus, $\widehat R_\ell$ is regular at the horizon. 

To sum up, due to the analytic structures of the two antiderivatives $A_{\hat u}$ and $E$ that we discussed above, each mode $\widehat R_\ell$ is a linear combination of
\begin{equation} \label{sm:function-class}
    1\,,\quad L\,,\quad H\,,\quad G\,,\quad
    \Phi_{z_+}\,,\quad \Phi_{z_-}\,,
\end{equation}
with rational coefficients in $z$. All functions are real-analytic outside the horizon $z>z_+$ on their principal branches. In particular, the dilogarithm arguments have negative real part there and never reach the cut $[1,\infty)$. Therefore, $R_\ell$ is closed form for each mode.

%%%%%%%%%%%%%%%%%%%%%%%%%%%%%%%%%%%%%%%%%%%%%%%%%%%%%%%%%%%%%%
%%%%%%%%%%%%%%%%%%%%%%%%%%%%%%%%%%%%%%%%%%%%%%%%%%%%%%%%%%%%%%
\section{Transport equations and metric reconstruction}
\label{smtc:transport}

%%%%%%%%%%%%%%%%%%%%%%%%%%%%%%%%%%%%%%%%%%%%%%%%%%%%%%%%%%%%%%
%%%%%%%%%%%%%%%%%%%%%%%%%%%%%%%%%%%%%%%%%%%%%%%%%%%%%%%%%%%%%%
\subsection{The integrating factors}
\label{sm:integrating-factors}

We use the traceful outgoing radiation gauge $h_{\mu\nu}^{(1)}n^\nu=0$ of Ref.~\cite{Li:2026rkf}, without imposing $g^{\mu\nu}h_{\mu\nu}^{(1)}=0$. The nonzero metric components are then $h_{ll}^{(1)}$, $h_{l\bar m}^{(1)}$, $h_{m\bar m}^{(1)}$, $h_{\bar m\bar m}^{(1)}$, and their complex conjugates. In this gauge, the linearized Newman--Penrose equations can be ordered so that each unknown obeys a first-order equation along $n^\mu$ whose source contains only quantities found earlier. Every such equation has the form $\Nop{n_1,n_2,n_3,n_4}f=S_f$. Since $\DD=-(\Delta/2\Sigma)\partial_r$ on stationary fields, it is a first-order ordinary differential equation in $r$, solved with the integrating factor 
\begin{equation} \label{smtc:integrating-factor}
    H_{[n_1,n_2,n_3,n_4]}
    =\frac{\Delta^{(n_3+n_4)/2}}{\Gb^{n_1+n_3}\Gm^{n_2+n_4}}\,,\qquad
    \frac{\partial_rH_{[n_1,n_2,n_3,n_4]}}{H_{[n_1,n_2,n_3,n_4]}}
    =\frac{2\Sigma}{\Delta}(n_1\mu+n_2\bar\mu+n_3\gamma+n_4\bar\gamma)\,,
\end{equation}
i.e., $\Nop{n_1,n_2,n_3,n_4}f=-(\Delta/2\Sigma)H_{[n_1,n_2,n_3,n_4]}\,\partial_r(f/H_{[n_1,n_2,n_3,n_4]})$. Integrating from infinity gives
\begin{equation} \label{smtc:tail-solution}
    f=H_{[n_1,n_2,n_3,n_4]}\left[c_f(\theta)+2\int_r^\infty
    \frac{\Sigma S_f}{\Delta H_{[n_1,n_2,n_3,n_4]}}\,d\tilde r\right]\,,
\end{equation}
where the first term involving the integration function $c_f(\theta)$ is a homogeneous solution. We will show in Sec.~\ref{smtc:completion} that all $c_f(\theta)$ vanish when proper conditions are imposed on the boundary and conserved charges of the spacetime.

%%%%%%%%%%%%%%%%%%%%%%%%%%%%%%%%%%%%%%%%%%%%%%%%%%%%%%%%%%%%%%
%%%%%%%%%%%%%%%%%%%%%%%%%%%%%%%%%%%%%%%%%%%%%%%%%%%%%%%%%%%%%%
\subsection{The transport hierarchy}
\label{sm:hierarchy}

The transport hierarchy of Ref.~\cite{Li:2026rkf} [its Eqs.~(13)--(24)] determines, in order, the perturbed spin coefficients $\mu^{(1)},\lambda^{(1)},\pi^{(1)}$, the perturbed Weyl scalars $\Psi_3^{(1)},\Psi_2^{(1)}$, and the four metric components. The perturbed Weyl scalars and spin coefficients need the $\mathcal{O}(\epsilon)$ correction to the tetrad. Here, we follow the convention in Ref.~\cite{Li:2026rkf} to pick
\begin{equation} \label{smtc:perturbed-tetrad}
    l^{\mu(1)}=\tfrac12h_{ll}^{(1)}n^\mu\,,\qquad 
    n^{\mu(1)}=0\,,\qquad
    m^{\mu(1)}=h_{lm}^{(1)}n^\mu-\tfrac12h_{m\bar m}^{(1)}m^\mu-\tfrac12h_{mm}^{(1)}\bar m^\mu\,.
\end{equation}
On the other hand, the Newman--Penrose Ricci scalars $\Phi_{ij}$, $i,j\in\{0,1,2\}$, and $\Lambda$ are evaluated on the Kinnersley tetrad of the Kerr metric since the latter is Ricci flat and the Ricci tensor at $\mathcal{O}(\epsilon^0)$ vanishes. As shown in Fig.~\ref{fig:flowchart}, the transport equations in our reconstruction hierarchy form two branches. The trace branch, $\Phi_{22}^{(1)}\to\mu^{(1)}\to h_{m\bar m}^{(1)}$, is driven only by the Ricci source. The $\Psi_4^{(1)}$ branch, $\Psi_4^{(1)}\to\lambda^{(1)}\to h_{\bar m\bar m}^{(1)}\to\Psi_3^{(1)}\to\pi^{(1)}\to h_{l\bar m}^{(1)} \to\Psi_2^{(1)}\to h_{ll}^{(1)}$, gives the remaining components, in the sequence of spin weight -2, -1, and 0, respectively. With the Ricci source terms of Ref.~\cite{Li:2026rkf} [its Eqs.~(30g) and (30h)] evaluated on the Kerr metric,
\begin{equation} \label{smtc:ricci-combinations}
    R_h^{(1)}
    =\Nop{0,2,2,0}\Phi_{21}^{(1)}
    -\dbop{2,2,0,-1}\Phi_{22}^{(1)}\,,\qquad
    R_g^{(1)}
    =\dop{0,2,2,0}\Phi_{21}^{(1)}
    -\Dop{2,2,0,-1}\Phi_{22}^{(1)}-2\DD\Lambda^{(1)}
    -2\mu\Phi_{11}^{(1)}+2\pi\Phi_{12}^{(1)}\,,
\end{equation}
the transport equations in the reconstruction hierarchy are
\begin{align} \label{smtc:hierarchy}
    & \Nop{2,0,1,1}\mu^{(1)}=-\Phi_{22}^{(1)}\,,
    && \Nop{-1,1,0,0}h_{m\bar m}^{(1)}=2\mu^{(1)}\,, \nonumber\\
    & \Nop{1,1,3,-1}\lambda^{(1)}
    =-\Psi_4^{(1)}\,,
    && \Nop{-1,1,2,-2}h_{\bar m\bar m}^{(1)}
    =2\lambda^{(1)}\,, \nonumber\\
    & \Nop{4,0,2,0}\Psi_3^{(1)}
    =\dop{0,4,0,-1}\Psi_4^{(1)}+R_h^{(1)}\,,
    && \Nop{0,0,1,-1}\pi^{(1)}
    =S^{(1)}_\pi\,, \nonumber\\
    & \Nop{0,1,0,-2}h_{l\bar m}^{(1)}
    =2\pi^{(1)}-\bar\tau h_{m\bar m}^{(1)}-\tau h_{\bar m\bar m}^{(1)}\,, 
    && \Nop{3,0,0,0}\Psi_2^{(1)}
    =-3\Psi_2\mu^{(1)}+\dop{0,2,0,-2}\Psi_3^{(1)}+R_g^{(1)}\,, \nonumber\\
    & \Nop{-1,1,-2,-2}h_{ll}^{(1)}
    =\frac{2}{\mu}\,S^{(1)}_\mu-B_h^{(1)}\,,
\end{align}   
where 
\begin{align} 
    & S^{(1)}_\pi
    =-\Psi_3^{(1)}-\Phi_{21}^{(1)}
    -\tfrac12(\pi+\bar\tau)\Nop{0,1,0,0}h_{m\bar m}^{(1)}
    -\tfrac12(\bar\pi+\tau)\Nop{0,1,2,-2}h_{\bar m\bar m}^{(1)}
    +\tfrac12(\mu-\bar\mu)\pi h_{m\bar m}^{(1)}\,, \nonumber\\
    & S^{(1)}_\mu
    =-\Dop{1,1,0,-1}\mu^{(1)}+\delta^{(1)}_{[-1,1,1,0]}\pi
    +\dop{-1,1,1,0}\pi^{(1)}+\Psi_2^{(1)}+2\Lambda^{(1)}\,, \nonumber\\
    & B_h^{(1)}
    =-\dbop{-2,0,-1,0}h_{lm}^{(1)}
    +\dop{-2,0,1,2}h_{l\bar m}^{(1)}
    +\Dop{0,0,-1,1}h_{m\bar m}^{(1)}\,,\nonumber\\
    & \alpha^{(1)}
    =\tfrac14\left(\dop{-2,0,1,1}h_{\bar m\bar m}^{(1)}
    +\Nop{-2,1,4,-2}h_{l\bar m}^{(1)}
    -\dbop{2,0,-1,-1}h_{m\bar m}^{(1)}\right)\,, \nonumber\\
    & \beta^{(1)}
    =\tfrac14\left(-\dbop{0,2,-1,-1}h_{mm}^{(1)}
    +\Nop{1,2,2,0}h_{lm}^{(1)}
    +\dop{0,-2,1,1}h_{m\bar m}^{(1)}\right)\,,\nonumber\\
    & \delta^{(1)}_{[-1,1,1,0]}\pi
    =h_{lm}^{(1)}\DD\pi-\tfrac12h_{m\bar m}^{(1)}\delta\pi
    -\tfrac12h_{mm}^{(1)}\bar\delta\pi
    +\left(-\bar\alpha^{(1)}+\beta^{(1)}+\bar\pi^{(1)}\right)\pi\,, 
\end{align}
and the sequence of reconstruction is from left to right and from top to bottom. Here, $S^{(1)}_\mu$ is the right-hand side of Eq.~(22) of Ref.~\cite{Li:2026rkf}. Since $\mu$ vanishes at the horizon, we do not divide by it and use the $h_{ll}^{(1)}$ equation in the form of Eqs.~(22) and~(23) of Ref.~\cite{Li:2026rkf} instead,
\begin{equation} \label{smtc:hll-undivided}
    \tfrac12\mu\left(\Nop{-1,1,-2,-2}h_{ll}^{(1)}+B_h^{(1)}\right)
    =S^{(1)}_\mu\,.
\end{equation}
The integrating factor $H$, defined in Eq.~\eqref{smtc:integrating-factor}, of each transport equation in Eq.~\eqref{smtc:hierarchy} (from left to right and from top to bottom) is 
\begin{equation} \label{smtc:integrating_factors_list}
    \frac{\Delta}{\Gm\Gb^3}\,,\quad
    \frac{\Gb}{\Gm}\,,\quad
    \frac{\Delta}{\Gb^4}\,,\quad
    \frac{\Gm}{\Gb}\,,\quad
    \frac{\Delta}{\Gb^6}\,,\quad
    \frac{\Gm}{\Gb}\,,\quad
    \frac{\Gm}{\Delta}\,,\quad
    \frac{1}{\Gb^3}\,,\quad
    \frac{\Gm\Gb^3}{\Delta^2}\,,
\end{equation}
respectively.

%%%%%%%%%%%%%%%%%%%%%%%%%%%%%%%%%%%%%%%%%%%%%%%%%%%%%%%%%%%%%%
%%%%%%%%%%%%%%%%%%%%%%%%%%%%%%%%%%%%%%%%%%%%%%%%%%%%%%%%%%%%%%
\subsection{Solving the transport equations}
\label{sm:moments}

The transport equations to get $h_{m\bar{m}}^{(1)}$ in Eq.~\eqref{smtc:hierarchy} are easy to solve. Inserting $\Phi_{22}^{(1)}$ from Eq.~\eqref{smtc:np-source}, we reduce the equation of $\mu^{(1)}$ to
\begin{equation}
 \partial_r\left(\frac{\Gm\Gb^3\mu^{(1)}}{\Delta}\right)
   =162M^2(\Gb^{-6}+\Gb^2\Gm^{-8}),\qquad
 \label{smtc:trace-antiderivative}
\end{equation}
so the solution to $\mu^{(1)}$ is 
\begin{equation}
    \mu^{(1)}
    =-162M^2\frac{\Delta}{\Gm\Gb^3}
    \left(\frac1{5\Gb^5}+\frac1{105\Gm^5}
    +\frac{\Gb}{21\Gm^6}+\frac{\Gb^2}{7\Gm^7}\right)\,.
\end{equation}
Then, the equation for $h_{m\bar{m}}^{(1)}$,
\begin{equation}
    \partial_r\left(\Gm h_{m\bar m}^{(1)}/\Gb\right)
    =-4\Gm^2\mu^{(1)}/\Delta\,,
\end{equation}
yields the solution for $h_{m\bar{m}}^{(1)}$ in Eq.~\eqref{smtc:hmmc} below. 

On the other hand, the branch driven by $\Psi_4^{(1)}$ is a bit more involved. Let us take $h_{\bar{m}\bar{m}}^{(1)}$ as an example. With Eq.~\eqref{smtc:reflection}, we can replace $\Psi_4^{(1)}$ with $\Psi_0^{(1)}$ in the transport equations of Eq.~\eqref{smtc:hierarchy}, so
\begin{equation} \label{smtc:shear-antiderivative}
    \partial_r\left(\frac{\Gb^4\lambda^{(1)}}{\Delta}\right)
    =\frac{\Sigma}{2}\Psi_0^{(1)}\,,\qquad 
    \partial_r\left(\frac{\Gb}{\Gm}h_{\bar m\bar m}^{(1)}\right)
    =-\frac{4\Gb^2}{\Delta}\lambda^{(1)}\,,
\end{equation}
and the solution for $h_{\bar{m}\bar{m}}^{(1)}$ takes the form
\begin{equation} \label{smtc:hmcmc_psi0}
    h_{\bar m\bar m}^{(1)}
    =-\frac{2\Gm}{\Gb^2}\int_r^\infty(\tilde r-r)
    (\tilde r+iax)\Psi_0^{(1)}(\tilde r,\theta)\,d\tilde r\,,
\end{equation}
where we have used Cauchy's formula for repeated integration to collapse the nested integrals in $r$, i.e.,
\begin{equation} \label{sm:cauchy}
    \int_r^\infty dr_1\int_{r_1}^\infty dr_2
    \cdots\int_{r_{n-1}}^\infty dr_n\,f(r_n)
    =\int_r^\infty\frac{(\tilde r-r)^{n-1}}{(n-1)!}
    f(\tilde r)\,d\tilde r\,.
\end{equation}
This motivates us to define the following moments $\mathcal{I}_n$,
\begin{equation} \label{smtc:moments}
    \mathcal I_n
    =\frac1{n!}\int_r^\infty(\tilde r-r)^n
    \Psi_0^{(1)}(\tilde r,\theta)\,d\tilde r\,,
\end{equation}
such that we can rewrite $\tilde r+iax=(\tilde r-r)+\Gm$ in
Eq.~\eqref{smtc:hmcmc_psi0} and obtain
\begin{equation}
    h_{\bar m\bar m}^{(1)}
    =-(2\Gm/\Gb^2)(2\mathcal I_2+\Gm\,\mathcal I_1)\,.
\end{equation}

The $h_{\bar{m}\bar{m}}^{(1)}$ example shows that the radial weight in the quadrature form of each transport equation's solution controls the moment order of $\mathcal{I}_n$. Two radial integrations give $h_{\bar m\bar m}^{(1)}$, but its quadratic radial kernel, i.e., $(\tilde{r}-r)^2$, requires $\mathcal I_2$. Similarly, the Weyl-driven part of the $\Psi_3^{(1)}$ transport equation contains a cubic weight $(\tilde r-iax)^3=[\Gb+(\tilde r-r)]^3$. With $\mathscr L_s=\partial_\theta+s\cot\theta$, its contribution includes
\begin{equation}
    \int_r^\infty(\tilde r-iax)^3
    \mathscr L_2\Psi_0^{(1)}(\tilde r,\theta)\,d\tilde r
    =\Gb^3\mathscr L_2\mathcal I_0
    +3\Gb^2\mathscr L_2\mathcal I_1
    +6\Gb\mathscr L_2\mathcal I_2
    +6\mathscr L_2\mathcal I_3\,.
\end{equation}
Explicit reduction of the subsequent equations for $\pi^{(1)}$, $h_{l\bar m}^{(1)}$, $\Psi_2^{(1)}$, and $h_{ll}^{(1)}$ closes on $\mathcal I_0,\ldots,\mathcal I_3$ and their angular derivatives, together with rational Ricci-source terms. This closure is also manifest in the metric of Eq.~\eqref{eq:metric_final}: the Weyl-driven parts of $h_{\bar m\bar m}^{(1)}$ and $h_{l\bar m}^{(1)}$ involve only $\{\mathcal I_1,\mathcal I_2\}$ and $\{\mathcal I_2,\mathcal I_3\}$, respectively, while the final component satisfies
\begin{equation}
    \frac{\Delta^2}{\Sigma^2}h_{ll}^{(1)}
    =-8\operatorname{Re}\left[
    \frac{1}{\Gb^2}
    \left(\mathscr L_1+\frac{2j}{\Gm}\right)
    \mathscr L_2\mathcal I_3\right]+J_R\,,
\end{equation}
where $j=ia\sin\theta$ and $J_R$ is the rational Ricci-source term in Eq.~\eqref{eq:correctors}. As the trace $h_{m\bar{m}}^{(1)}$ is also determined entirely by the Ricci source, no moment beyond $\mathcal I_3$ is required. 

We therefore introduce the single potential $\Pot\equiv\mathcal I_3$. Differentiating the moment definition gives $\partial_r\mathcal I_n=-\mathcal I_{n-1}$ for $n\ge1$ and $\partial_r\mathcal I_0=-\Psi_0^{(1)}$, so
\begin{equation}
    \mathcal I_2=-\partial_r\Pot\,,\qquad
    \mathcal I_1=\partial_r^2\Pot\,,\qquad
    \mathcal I_0=-\partial_r^3\Pot\,,\qquad
    \Psi_0^{(1)}=\partial_r^4\Pot\,.
    \label{smtc:moment-identities}
\end{equation}
All Weyl-driven metric terms can consequently be written as derivatives of $\Pot$, with the rational Ricci-source terms retained separately. For a given $\Psi_0^{(1)}$, the integral definition fixes the four integration functions of $\theta$ left undetermined by $\partial_r^4\Pot=\Psi_0^{(1)}$. Homogeneous solutions of the transport equations are treated separately in Sec.~\ref{smtc:completion}, where they are set to zero by fixing ADM charges and imposing asymptotic conditions for the metric in the orthonormal frame.

%%%%%%%%%%%%%%%%%%%%%%%%%%%%%%%%%%%%%%%%%%%%%%%%%%%%%%%%%%%%%%
%%%%%%%%%%%%%%%%%%%%%%%%%%%%%%%%%%%%%%%%%%%%%%%%%%%%%%%%%%%%%%
\subsection{The potential equation and its integral-free solution}
\label{sec:supp-source-moment}

Although computing $\mathcal P$ from its definition would require three more radial integrals of $\Psi_0^{(1)}$, we avoid them by using the fact that $\mathcal P$ obeys a stationary Teukolsky equation with a rational source, which is related to the stationary Teukolsky equation of $R_\ell$, so $\mathcal P$ follows from $R_\ell$ by differentiation.

With $\mathcal D_+=\Delta\partial_r^2+6(r-M)\partial_r$ and
$\mathcal D_-=\Delta\partial_r^2-2(r-M)\partial_r$, direct differentiation gives the operator identities 
\begin{equation}
 \Delta^2\partial_r^4=\mathcal D_-(\mathcal D_-+2),\qquad
 \partial_r^4\mathcal D_-=(\mathcal D_++4)\partial_r^4 ,
 \label{sm:ts-identities}
\end{equation}
which hold for arbitrary functions. 

The operators $\mathcal H_0$ and $\mathcal H_4$ of Eq.~\eqref{sm:teuk-operators} share their angular part and differ by $\mathcal H_0-\mathcal H_4=\mathcal D_+-\mathcal D_-+4$, so the second identity of Eq.~\eqref{sm:ts-identities} gives $\partial_r^4\mathcal H_4=\mathcal H_0\partial_r^4$. Thus, $\partial_r^4(\mathcal H_4\mathcal P)=\mathcal H_0\partial_r^4\mathcal P=\mathcal H_0\Psi_0^{(1)}=\mathcal T_2$. The Teukolsky
source $\mathcal{T}_2$ [see Eq.~\eqref{sm:teuk-source}] has an elementary fourth radial antiderivative that vanishes at infinity:
\begin{equation} \label{eq:supp-full-source-moment}
    Q_{\rm full}
    =-\frac{20}{7}M^2a^2(1-x^2)\frac{5r-iax}{\Gm^7}\,,\qquad
    \partial_r^4 Q_{\rm full}
    =\mathcal T_2\,.
\end{equation}
Hence, $\partial_r^4(\mathcal H_4\mathcal P-Q_{\rm full})=0$. Because both $\mathcal P$ and $Q_{\rm full}$ decay at infinity, the cubic polynomial in $r$ allowed by $\partial_r^4$ is excluded, so the potential $\Pot$ obeys the stationary spin-$(-2)$ Teukolsky equation with a rational source: 
\begin{equation}
    \mathcal H_4\mathcal P=Q_{\rm full}.
\end{equation}

The potential $\Pot$ separates in the same basis as $\Psi_0^{(1)}$. The angular part of $\mathcal H_4$ in Eq.~\eqref{sm:teuk-operators} is
$(1-x^2)\partial_x^2-2x\partial_x-4/(1-x^2)+2$, where the first three terms form the associated Legendre operator with $m=2$, whose eigenvalue on $P_\ell^2$ is $-\ell(\ell+1)$. Hence
\begin{equation} \label{eq:supp-H4-mode}
    \mathcal H_4\big(I_\ell P_\ell^2\big)
    =\big(\mathcal D_--k_\ell\big)I_\ell\,P_\ell^2\,,\qquad 
    k_\ell=\ell(\ell+1)-2=(\ell-1)(\ell+2)\,.
\end{equation}
With $\mathcal P=\sum_\ell I_\ell P_\ell^2$, projecting $Q_{\rm full}$ onto $P_\ell^2$ with the norm of Eq.~\eqref{smtc:radial-equation} gives the source in the equation of $\mathcal{P}$:
\begin{equation} \label{eq:supp-source-projections}
\nu_\ell\int_{-1}^{1}P_\ell^2\, Q_{\rm full}\,dx
    =\frac16\int_r^\infty(\tilde r-r)^3T_\ell\,d\tilde r
    \equiv\mathcal F_\ell\,,
\end{equation}
where the second integral follows from
$Q_{\rm full}=\frac16\int_r^\infty(\tilde r-r)^3\mathcal T_2\,d\tilde r$ (Cauchy's formula~\eqref{sm:cauchy} for the fourth antiderivative that vanishes at infinity) and from exchanging the order of the $x$ and $\tilde r$ integrals. Therefore, the equation of $I_\ell$ is
\begin{equation} \label{eq:supp-radial-potential}
    \big(\mathcal D_--k_\ell\big)I_\ell =\mathcal F_\ell\,.
\end{equation}
Since $\partial_r^4I_\ell=R_\ell$, the first identity in Eq.~\eqref{sm:ts-identities} gives $\Delta^2R_\ell=\mathcal D_-(\mathcal D_-+2)I_\ell$. Using the equation $\mathcal D_-I_\ell=k_\ell I_\ell+\mathcal F_\ell$ twice and $k_\ell(k_\ell+2)=K_\ell$, we find $\Delta^2R_\ell=K_\ell I_\ell+\mathcal E_\ell$ with $\mathcal E_\ell=(\mathcal D_-+k_\ell+2)\mathcal F_\ell$, so
\begin{equation} \label{sm:adjoint-physical}
    I_\ell=\frac{\Delta^2R_\ell-\mathcal E_\ell}{K_\ell}\,.
\end{equation}
Hence, no integration of $R_\ell$ is needed to obtain $I_{\ell}$, the latter of which is the radial part of the potential $\mathcal{P}$. 

Now, we examine the analytic structure of $\mathcal E_\ell$ by integrating the source in Eq.~\eqref{eq:supp-source-projections}. In the variables of Sec.~\ref{sm:integral-free}, the source is $\mathcal F_\ell(r)=a^4\mathcal N_\ell\widehat{\mathcal F}_\ell(z)$ with
\begin{equation} \label{sm:F-hat}
    \widehat{\mathcal F}_\ell(z)
    =\frac16\int_z^\infty(\tilde z-z)^3
    J_\ell(\tilde z)\,d\tilde z\,,
\end{equation}
where the hat variables are the dimensionless counterparts of the original ones. Expanding $(\tilde z-z)^3=\tilde z^3-3z\tilde z^2+3z^2\tilde z-z^3$ reduces $\widehat{\mathcal F}_\ell(z)$ to four single integrals,
\begin{equation}
    \mathfrak t_n(z)
    =\int_z^\infty\tilde z^{\,n}J_\ell(\tilde z)\,d\tilde z\,,
    \qquad n=0,1,2,3\,,
\end{equation}
where $\mathfrak t_n(z)$ contains only rational functions and a polynomial times $L$.  The modes of the potential $\mathcal P$ are then
\begin{equation} \label{sm:potential-closed}
    \widehat I_\ell
    =\frac{\hat\Delta^2\widehat R_\ell-\widehat{\mathcal E}_\ell}{K_\ell}\,,
    \qquad
    \widehat{\mathcal E}_\ell
    =(k_\ell+2)\widehat{\mathcal F}_\ell
    -2(z-\hat M)\widehat{\mathcal F}_\ell'
    +\hat\Delta\widehat{\mathcal F}_\ell''\,,\qquad
    \widehat{\mathcal F}_\ell
    =\frac{\mathfrak t_3-3z\mathfrak t_2
    +3z^2\mathfrak t_1-z^3\mathfrak t_0}{6}\,,
\end{equation}
while $I_\ell(r)=a^4\mathcal N_\ell\widehat I_\ell(z)$. Since $\widehat{\mathcal F}_\ell$ are only rational functions and a polynomial times $L$, $\widehat{\mathcal E}_\ell$ contributes no additional algebraic structure [see Eq.~\eqref{sm:function-class}] to $\widehat R_\ell$. Thus, $\widehat I_\ell$ contains the same group of functions as $\widehat R_\ell$.

%%%%%%%%%%%%%%%%%%%%%%%%%%%%%%%%%%%%%%%%%%%%%%%%%%%%%%%%%%%%%%
%%%%%%%%%%%%%%%%%%%%%%%%%%%%%%%%%%%%%%%%%%%%%%%%%%%%%%%%%%%%%%
\subsection{Explicit metric}
\label{smtc:explicit-metric}

Integrating the hierarchy in Eq.~\eqref{smtc:hierarchy} from infinity (all $c_f=0$, Sec.~\ref{smtc:completion}) and using Eq.~\eqref{smtc:moment-identities}, every $\Psi_4^{(1)}$ contribution becomes a derivative of $\Pot$. 
With the angular operator $\mathscr{L}_k=\partial_\theta+k\cot\theta$ ($k$ is an integer) and $S_{[m,n]}=\Gm^{-m}\Gb^{-n}+\Gm^{-n}\Gb^{-m}$, the four independent metric components in our reconstruction hierarchy are
\begin{subequations}
\begin{align}
    & h_{m\bar m}^{(1)}
    =-\frac{108M^2}{35}\left[6(\Gm^{-6}+\Gb^{-6})+S_{[1,5]}\right]\,,
    \label{smtc:hmmc} \\
    & h_{\bar m\bar m}^{(1)}
    =-\frac{2\Gm}{\Gb^2}(\Gm\partial_r^2-2\partial_r)\mathcal{P}\,,
    \label{smtc:hmcmc} \\
    & h_{l\bar m}^{(1)}
    =\frac{2\sqrt2\,\Gm}{\Gb^2\Delta}
    \left[\Sigma\mathscr{L}_2\partial_r-(\Gm+\Gb)\mathscr{L}_2
    -j(\Gm-\Gb)\partial_r\right]\mathcal{P}
    +\frac{\sqrt2j\,\Gm}{\Delta}G_R\,,
    \label{smtc:hlmc} \\
    & h_{ll}^{(1)}
    =\frac{\Sigma^2}{\Delta^2}\left[-8\operatorname{Re}
    \left\{\frac1{\Gb^2}\left(\mathscr{L}_1+\frac{2j}{\Gm}\right)
    \mathscr{L}_2\mathcal{P}\right\}+J_R\right]\,,
    \label{smtc:hll}
\end{align}    
\end{subequations}
where
\begin{align}
    & G_R=\frac{M^2}{35}\left(
    \frac{368}{\Gm^6}+\frac{348}{\Gm^5\Gb}+\frac{40}{\Gm^4\Gb^2}
    +\frac{108}{\Gm\Gb^5}+\frac{648}{\Gb^6}\right)\,, \nonumber\\
    & J_R=\frac{16M^3}{35}(-8S_{[7,0]}-5S_{[6,1]}+27S_{[5,2]})
    -\frac{16M^2a^2}{35}
    \left[92s^2S_{[7,1]}+(87-67x^2)S_{[6,2]}\right]\,.
\end{align}
Each metric component $h_{ab}^{(1)}$ is thus a differential operator of at most second order acting on $\Pot$, plus rational functions.
The trace $h_{m\bar m}^{(1)}$ and the rational functions, including $G_R$ and $J_R$, come from the Ricci source, while remaining terms involving $\Pot$ come from $\Psi_4^{(1)}$. The factors $\Delta^{-1}$ and $\Delta^{-2}$ in $h_{l\bar m}^{(1)}$ and $h_{ll}^{(1)}$ only reflect the normalization of the Kinnersley vector $l^\mu$, which diverges at the horizon. The combinations
\begin{equation} \label{smtc:regular-fields}
    C\equiv\frac{\Delta}{\Sigma}h_{l\bar m}^{(1)}\,,\qquad
    \mathcal{J}\equiv\frac{\Delta^2}{\Sigma^2}h_{ll}^{(1)}
\end{equation}
have only $\Gm$ and $\Gb$ in their denominators. In terms of $\mathcal J$, the $h_{ll}^{(1)}$ equation reads
\begin{equation} \label{smtc:J-transport}
    \partial_r\mathcal{J}
    =\frac{2iax}{\Sigma}\mathcal{J}+\boldsymbol{\mathcal{F}}\,,\qquad
    \boldsymbol{\mathcal{F}}
    =8\Gb S^{(1)}_\mu+\frac{2\Delta}{\Sigma}B_h^{(1)}\,.
\end{equation}
Since $\mathcal J$ is real while this equation is complex, this imposes two conditions, a radial equation and an algebraic constraint:
\begin{equation} \label{smtc:reality}
    \partial_r\mathcal{J}=\operatorname{Re}\boldsymbol{\mathcal{F}},\qquad
    \frac{2ax}{\Sigma}\mathcal{J}+\operatorname{Im}\boldsymbol{\mathcal{F}}=0\,,
\end{equation}
where the first equation was used in our reconstruction hierarchy, while the second equation is imposed by the Einstein-equation closure that we will prove in Sec.~\ref{smtc:closure}.

The ingoing coordinates $(v,r,\theta,\varphi)$ are related to the Boyer--Lindquist coordinates $(t,r,\theta,\phi)$ by
\begin{equation} \label{smtc:bl-transformation}
    dv=dt+\frac{r^2+a^2}{\Delta}\,dr,\qquad
    d\varphi=d\phi+\frac{a}{\Delta}\,dr\,.
\end{equation} 
With $\varpi^2=r^2+a^2$, $c_0=\operatorname{Re}\left[C/\Gm\right]$, $c_1=\operatorname{Re}\left[iC/\Gm\right]$, $b_0=\operatorname{Re}\left[h_{\bar m\bar m}^{(1)}/\Gm^2\right]$, and
$b_1=\operatorname{Re}\left[ih_{\bar m\bar m}^{(1)}/\Gm^2\right]$, the nonzero components in the ingoing coordinates are 
\begin{align} \label{smtc:coordinate-components}
    & h_{vv}^{(1)}=\frac{\mathcal{J}}4-\sqrt2as c_1-a^2s^2b_0+\frac{a^2s^2}{\Sigma}h_{m\bar m}^{(1)}\,,\quad
    h_{v\theta}^{(1)}=\frac{\Sigma}{\sqrt2}c_0-as\Sigma b_1\,, \nonumber\\
    & h_{v\varphi}^{(1)}
    =-\frac{as^2\mathcal{J}}4+\frac{s(\varpi^2+a^2s^2)}{\sqrt2}c_1
    +a\varpi^2s^2b_0-\frac{a\varpi^2s^2}{\Sigma}h_{m\bar m}^{(1)}\,, \nonumber\\
    & h_{\theta\theta}^{(1)}
    =\Sigma^2b_0+\Sigma h_{m\bar m}^{(1)}\,,\quad
    h_{\theta\varphi}^{(1)}
    =-\frac{as^2\Sigma}{\sqrt2}c_0+\varpi^2s\Sigma b_1\,,\quad
    h_{\varphi\varphi}^{(1)}
    =\frac{a^2s^4\mathcal{J}}4-\sqrt2a\varpi^2s^3c_1
    -\varpi^4s^2b_0+\frac{\varpi^4s^2}{\Sigma}h_{m\bar m}^{(1)}\,.
\end{align}

$\mathcal J$ and $\mathcal C$ contain no factor $\Delta^{-1}$, and the coefficients $h_{m\bar m}^{(1)}$, $h_{\bar m\bar m}^{(1)}$, $C$, and $\mathcal J$ have only $\Gm$ and $\Gb$ in their denominators, which do not vanish for $r\geq r_+$. The perturbation in Eq.~\eqref{smtc:coordinate-components} is therefore regular at the future horizon in the ingoing chart. At the axes, $h_{\bar m\bar m}^{(1)}=\mathcal{O}(\sin^2\theta)$ and $C=\mathcal{O}(\sin\theta)$, while $h_{m\bar m}^{(1)}$ and $\mathcal J$ are smooth in $\cos\theta$, so $h_{\mu\nu}^{(1)}$ is regular there as well. Given that $\delta g^{rr}=-\epsilon \mathcal J/4$, the horizon position of our black-hole solution in parity-even cubic gravity gets displaced by $\delta r_{+}$, where
\begin{equation}  \label{smtc:horizon-shift}
    \delta r_{+}=\frac{\Sigma_+\mathcal J_+}{4\Delta'_+}\,,
\end{equation}
and the subscript $+$ denotes evaluation at the Kerr horizon $r=r_+$.

%%%%%%%%%%%%%%%%%%%%%%%%%%%%%%%%%%%%%%%%%%%%%%%%%%%%%%%%%%%%%%
%%%%%%%%%%%%%%%%%%%%%%%%%%%%%%%%%%%%%%%%%%%%%%%%%%%%%%%%%%%%%%
\subsection{Metric in Boyer--Lindquist coordinates}
\label{smtc:bl-metric}
 
The background is the Kerr metric in its usual form,
\begin{equation} \label{smtc:kerr-bl}
    ds_{(0)}^2
    =-\Big(1-\frac{2Mr}{\Sigma}\Big)dt^2-\frac{4Mar s^2}{\Sigma}\,dt\,d\phi
    +\frac{\Sigma}{\Delta}\,dr^2+\Sigma\,d\theta^2
    +\Big(\varpi^2+\frac{2Ma^2rs^2}{\Sigma}\Big)s^2d\phi^2\,.
\end{equation}
Using Eq.~\eqref{smtc:bl-transformation} to transform the metric to the Boyer--Lindquist coordinates, with the symbols defined before Eq.~\eqref{smtc:coordinate-components}, the nonzero components of $h_{\mu\nu}^{(1)}$ are
\begin{align}
    & h_{tt}^{(1)}
    =\frac{\mathcal J}{4}-\sqrt2\,as\,c_1-a^2s^2b_0+\frac{a^2s^2}{\Sigma}h_{m\bar m}^{(1)}\,,\qquad
    h_{tr}^{(1)}
    =\frac{\Sigma}{\Delta}\Big(\frac{\mathcal J}{4}
    -\frac{as}{\sqrt2}\,c_1\Big)\,,\qquad
    h_{t\theta}^{(1)}
    =\frac{\Sigma}{\sqrt2}\,c_0-as\Sigma\, b_1\,,\nonumber\\
    & h_{t\phi}^{(1)}
    =-\frac{as^2\mathcal J}{4}+\frac{s(\varpi^2+a^2s^2)}{\sqrt2}c_1
    +a\varpi^2s^2b_0-\frac{a\varpi^2s^2}{\Sigma}
    h_{m\bar m}^{(1)}\,,\nonumber\\
    & h_{rr}^{(1)}
    =\frac{\Sigma^2}{4\Delta^2}\,\mathcal J
    =\frac14h_{ll}^{(1)}\,,\qquad
    h_{r\theta}^{(1)}
    =\frac{\Sigma^2}{\sqrt2\,\Delta}\,c_0\,,\qquad
    h_{r\phi}^{(1)}
    =\frac{\Sigma}{\Delta}\Big(-\frac{as^2\mathcal J}{4}+\frac{\varpi^2s}{\sqrt2}\,c_1\Big)\,,\nonumber\\
    & h_{\theta\theta}^{(1)}
    =\Sigma^2b_0+\Sigma h_{m\bar m}^{(1)}\,,\qquad
    h_{\theta\phi}^{(1)}
    =-\frac{as^2\Sigma}{\sqrt2}\,c_0+\varpi^2s\Sigma\,b_1\,,\qquad
    h_{\phi\phi}^{(1)}
    =\frac{a^2s^4\mathcal J}{4}-\sqrt2\,a\varpi^2s^3c_1
    -\varpi^4s^2b_0+\frac{\varpi^4s^2}{\Sigma}h_{m\bar m}^{(1)}\,.
 \label{smtc:bl-components}
\end{align}
The factors $\Delta^{-1}$ and $\Delta^{-2}$ in these components are the usual coordinate singularities of Boyer--Lindquist coordinates at the horizon, as in the Kerr component $g^{(0)}_{rr}=\Sigma/\Delta$. As shown earlier in Sec.~\ref{smtc:explicit-metric}, the perturbation is regular in the ingoing chart.

%%%%%%%%%%%%%%%%%%%%%%%%%%%%%%%%%%%%%%%%%%%%%%%%%%%%%%%%%%%%%%
%%%%%%%%%%%%%%%%%%%%%%%%%%%%%%%%%%%%%%%%%%%%%%%%%%%%%%%%%%%%%%
\subsection{Nonrotating limit}
\label{sm:static}

At $a=0$, the Teukolsky source vanishes, so $\Psi_0^{(1)}=0$ and the potential terms drop out, but the rational Ricci terms do not. Thus, the surviving metric components are
\begin{equation} \label{smtc:static-org}
    h_{m\bar m}^{(1)}=-\frac{216M^2}{5r^6},\qquad h_{\bar m\bar m}^{(1)}=h_{l\bar m}^{(1)}=0\,,\qquad
    h_{ll}^{(1)}=\frac{r^4}{\Delta^2}\,\frac{64M^3}{5r^7}\,.
\end{equation}
In terms of the areal radius $R=r-108\epsilon M^2/(5r^5)$ and the static time $t$, the full line element at $a=0$ is
\begin{align}  \label{smtc:static-areal}
    & ds^2=-A_sdt^2+B_s^{-1}dR^2+R^2d\Omega^2\,, \nonumber\\
    & A_s=1-\frac{2M}{R}+\frac{40\epsilon M^3}{R^7},\qquad
    B_s=1-\frac{2M}{R}+\epsilon\left(\frac{216M^2}{R^6}-\frac{392M^3}{R^7}\right)\,.
\end{align}
This static solution agrees exactly with the $\chi^0$ part of the parity-even cubic solution of Ref.~\cite{Cano:2019ore} (with $\epsilon=\lambda_{\rm ev}\ell_{\rm EFT}^4$), once their radial coordinate is converted to the areal radius. It thus provides an independent check of our reconstruction in the nonrotating limit.

%%%%%%%%%%%%%%%%%%%%%%%%%%%%%%%%%%%%%%%%%%%%%%%%%%%%%%%%%%%%%%
%%%%%%%%%%%%%%%%%%%%%%%%%%%%%%%%%%%%%%%%%%%%%%%%%%%%%%%%%%%%%%
\section{Integration constants, charges, and gauge}
\label{smtc:completion}

In the Letter, we showed that horizon regularity and the absence of an external tidal field fix the homogeneous solutions to the radial Teukolsky equation. 
The Letter also gave the homogeneous transport solutions and the conditions that fix their integration functions. Here, we supply the asymptotic and charge-completion details, followed by a proof that the remaining Einstein equations hold.

%%%%%%%%%%%%%%%%%%%%%%%%%%%%%%%%%%%%%%%%%%%%%%%%%%%%%%%%%%%%%%
%%%%%%%%%%%%%%%%%%%%%%%%%%%%%%%%%%%%%%%%%%%%%%%%%%%%%%%%%%%%%%
\subsection{Asymptotic conditions and charge completion}
\label{sm:transport-data}

We use the asymptotically inertial, mass-centered representative specified in the Letter, with
\begin{equation} \label{smtc:falloff}
    h_{\hat a\hat b}^{(1)}=\mathcal O(r^{-3})\,,\qquad
    \nabla h^{(1)}=\mathcal O(r^{-4})\,,\qquad
    \nabla^2h^{(1)}=\mathcal O(r^{-5})\,,
\end{equation}
where all bounds refer to orthonormal-frame components of the metric perturbation and its derivatives. At leading order, the frame is $e_{\hat t}\sim\partial_t$, $e_{\hat r}\sim\partial_r$, $e_{\hat\theta}\sim r^{-1}\partial_\theta$, and $e_{\hat\varphi}\sim(r\sin\theta)^{-1}\partial_\phi$. The Kinnersley tetrad approaches $l=e_{\hat t}+e_{\hat r}$, $n=(e_{\hat t}-e_{\hat r})/2$, and $m=(e_{\hat\theta}+ie_{\hat\varphi})/\sqrt2$, so the Newman--Penrose metric components also satisfy $h_{ab}^{(1)}=\mathcal O(r^{-3})$.

The reconstructed particular solution has the required falloff. Its modes decay as $R_\ell=\mathcal O(r^{-5})$ or faster, given $\mathcal I_n=\mathcal O(r^{n-4})$ for $0\le n\le3$ and $\Pot=\mathcal O(r^{-1})$. Substitution into Eq.~\eqref{eq:metric_final} gives potential contributions of $\mathcal O(r^{-3})$, a trace component $h_{m\bar m}^{(1)}=\mathcal O(r^{-6})$, and remaining rational-source contributions of $\mathcal O(r^{-7})$. Their differentiated asymptotic expansions satisfy the derivative bounds in Eq.~\eqref{smtc:falloff}. 
By contrast, the homogeneous terms in Eq.~\eqref{eq:homog} of the Letter have successive leading orders no faster than $r^{-2}$. The main-text argument therefore sets every $c_f(\theta)=0$, including those multiplying decaying homogeneous solutions.

For homogeneous perturbations that also satisfy the remaining vacuum Einstein equations, Wald's theorem~\cite{Wald:1973wwa} identifies the non-gauge freedom with linearized mass, spin, etc. The mass and angular-momentum variations admit an explicit completion in the traceful outgoing radiation gauge we used above. 
Varying the Kerr background $g^{(0)}_{\mu\nu}$ at fixed ingoing coordinates $(v,r,x,\varphi)$ gives
\begin{align}
    & h^{\rm comp}_{\mu\nu}
    =\delta M\,\partial_Mg^{(0)}_{\mu\nu}\big|_a
    +\frac{\delta J_{\rm ADM}-a\delta M}{M}
    \left(\partial_ag^{(0)}_{\mu\nu}\big|_M
    +\mathcal{L}_\chi g^{(0)}_{\mu\nu}\right)\,,\nonumber\\
    & \chi
    =\left(-\frac{ar(1-x^2)}\Sigma,\,
    -\frac{ar(1-x^2)}\Sigma,\,
    \frac{ax(1-x^2)}\Sigma,\,-\frac r\Sigma\right)\,,
    \label{smtc:charge-completion}
\end{align}
where the components of $\chi$ are ordered as $(v,r,x,\varphi)$ and $(\delta J_{\rm ADM}-a\delta M)/M=\delta a$. The Lie derivative $\mathcal{L}_\chi g^{(0)}_{\mu\nu}$ restores $h^{\rm comp}_{r\mu}=0$, equivalent to the traceful outgoing radiation gauge in these coordinates, by canceling $\partial_ag^{(0)}_{r\varphi}=-(1-x^2)$ without changing either charge. 
Mass and angular-momentum shifts contribute at $\mathcal{O}(r^{-1})$ and $\mathcal{O}(r^{-2})$, respectively, so Eq.~\eqref{smtc:falloff} excludes them. Thus $\delta M=\delta J_{\rm ADM}=0$, and the reconstructed black hole has the same ADM mass and angular momentum as the Kerr background.

%%%%%%%%%%%%%%%%%%%%%%%%%%%%%%%%%%%%%%%%%%%%%%%%%%%%%%%%%%%%%%
%%%%%%%%%%%%%%%%%%%%%%%%%%%%%%%%%%%%%%%%%%%%%%%%%%%%%%%%%%%%%%
\subsection{Closure of the remaining Einstein equations}
\label{smtc:closure}

Our reconstruction hierarchy uses only part of the Newman--Penrose equations, so the remaining components of $\delta G_{\mu\nu}[h^{(1)}]=\mathcal S_{\mu\nu}$ must be checked. Let $P_{ij}=\Phi_{ij}^{(1)}[h^{(1)}]-\Phi_{ij}^{(1),\rm source}$ and $P_\Lambda=\Lambda^{(1)}[h^{(1)}]-\Lambda^{(1),\rm source}$ be the residuals of these equations. After the reconstruction, including the second equation in Eq.~\eqref{smtc:reality}, the remaining Ricci and Bianchi identities reduce to homogeneous equations for the residuals, i.e.,
\begin{align} \label{smtc:residual-equations}
    & P_{22}=0,\quad P_{11}=3P_\Lambda\,,\quad
    \Nop{2,1,2,0}P_{21}=0\,,\quad
    \Nop{1,1,0,0}P_\Lambda=0\,,\nonumber\\
    & \Nop{4,1,3,-1}\Nop{0,1,2,-2}P_{20}=0\,,\quad
    \Nop{1,2,-2,0}P_{01}=0\,, \quad
    \Nop{1,1,-2,-2}P_{00}=0\,,
\end{align}
with the general solutions
\begin{equation} \label{smtc:residual-modes}
    P_{21}=c_{21}\frac{\Delta}{\Gm\Gb^4}\,,\quad
    P_\Lambda=\frac{c_\Lambda}{\Sigma}\,,\quad
    P_{01}=c_{01}\frac{\Gb}{\Delta\Gm^2}\,,\quad
    P_{00}=c_{00}\frac{\Sigma}{\Delta^2}\,,\quad
    P_{20}=a_{20}\frac{\Gm}{\Gb^2}+b_{20}\frac{\Gm}{\Gb^5}\,.
\end{equation}
The fastest decay of the above residuals is from the last term, which decays as $r^{-4}$ at infinity. The falloff in Eq.~\eqref{smtc:falloff} makes the tetrad components of $\delta R_{\mu\nu}$ of $\mathcal{O}(r^{-5})$, and the cubic Ricci source is $\mathcal{O}(r^{-8})$, so all the coefficients in Eq.~\eqref{smtc:residual-modes} vanish, and the full linearized Einstein equation holds.

%%%%%%%%%%%%%%%%%%%%%%%%%%%%%%%%%%%%%%%%%%%%%%%%%%%%%%%%%%%%%%
%%%%%%%%%%%%%%%%%%%%%%%%%%%%%%%%%%%%%%%%%%%%%%%%%%%%%%%%%%%%%%
\section{Einstein residual and angular convergence rate}
\label{sec:supp-validation}

In practice, the mode sum for $\mathcal P$ is truncated at $\lmax$. Section~\ref{sec:finite-cutoff} derives the identity Eq.~\eqref{eq:identity} that appears in the main body of the Letter, which shows the residual of the Einstein equation due to a finite cutoff in the mode number of the solution to $\Psi_4^{(1)}$. We then examine the convergence rate of the residual in Sec.~\ref{sec: angular convergence}.

%%%%%%%%%%%%%%%%%%%%%%%%%%%%%%%%%%%%%%%%%%%%%%%%%%%%%%%%%%%%%%
%%%%%%%%%%%%%%%%%%%%%%%%%%%%%%%%%%%%%%%%%%%%%%%%%%%%%%%%%%%%%%
\subsection{Finite-cutoff Einstein identity} 
\label{sec:finite-cutoff}

We split the metric in Eqs.~\eqref{smtc:hmcmc}--\eqref{smtc:hll} into two parts, $h^{(1)}=h^{\rm pot}+h^{\rm rat}$. The potential part $h^{\rm pot}$ carries every term with $\mathcal P$, i.e., all of $h_{\bar m\bar m}^{(1)}$ and the $\mathcal P$ terms of $h_{l\bar m}^{(1)}$ and $h_{ll}^{(1)}$. $h^{\rm pot}$ carries the $\Psi_4^{(1)}$ content and is the only part affected by the truncation. The rational part $h^{\rm rat}$ consists of the Ricci-source terms, $h_{m\bar m}^{(1)}$ and the $G_R$ and $J_R$ terms, which are explicit rational functions of $\Gm$ and $\Gb$ and are kept exactly.

The linearized Einstein tensor is computed directly from the metric perturbation. On the Ricci-flat background,
\begin{equation}  \label{eq:supp-coordinate-einstein}
    \delta\Gamma^{\mu}{}_{\nu\rho}
    =\tfrac12 g^{\mu\sigma}
    (\nabla_\nu h_{\rho\sigma}^{(1)}
    +\nabla_\rho h_{\nu\sigma}^{(1)}
    -\nabla_\sigma h_{\nu\rho}^{(1)})\,,\quad
    \delta R_{\mu\nu}
    =\nabla_\rho\delta\Gamma^\rho{}_{\mu\nu}
    -\nabla_\nu\delta\Gamma^\rho{}_{\mu\rho}\,,\quad
    \delta G_{\mu\nu}
    =\delta R_{\mu\nu}
    -\tfrac12 g_{\mu\nu}g^{\rho\sigma}\delta R_{\rho\sigma}\,.
\end{equation}
To see how the truncation enters, we remove the angular factor $1-x^2$ from the potential and its source, $\mathcal P=(1-x^2)F$ and $Q_{\rm full}=(1-x^2)W$, so that
$F=\sum_\ell I_\ell P_\ell''$ and $W=\sum_\ell\mathcal F_\ell P_\ell''$. Since
$\mathcal H_4[(1-x^2)F]=(1-x^2)\mathscr QF$ with
\begin{equation} \label{eq:supp-adjoint-operator}
    \mathscr Q=\Delta\partial_r^2-2(r-M)\partial_r
    +(1-x^2)\partial_x^2-6x\partial_x-4\,,
\end{equation}
the equation $\mathcal H_4\mathcal P=Q_{\rm full}$ becomes $\mathscr QF=W$. Acting on each mode, $\mathscr Q(I_\ell P_\ell'')=[(\mathcal D_- -k_\ell)I_\ell]P_\ell''$, so Eq.~\eqref{eq:supp-radial-potential} is the mode-by-mode form of $\mathscr QF=W$. Truncating the potential at $\lmax$ therefore replaces the source by its truncated sum, $\mathscr QF_{\lmax}=W_{\lmax}$, with $W_{\lmax}=\sum_{\ell=2}^{\lmax}\mathcal F_\ell P_\ell''$.
 
The key property of the potential part is that its Einstein tensor depends on $F$ only through $\mathscr QF$. To show this, we keep $F$ arbitrary, i.e., we do not assume $\mathscr QF=W$, and evaluate Eq.~\eqref{eq:supp-coordinate-einstein} on $h^{\rm pot}$. Since the metric is real, $h^{\rm pot}$ contains both $F$ and $\bar F$, so we write $F=U+iV$ with real $U$ and $V$. Each of the ten components of $\delta G_{\mu\nu}[h^{\rm pot}]$ is then a linear combination, with rational coefficients, of the derivatives $\partial_r^i\partial_x^jU$ and $\partial_r^i\partial_x^jV$ with $i+j\leq4$. Because $\mathscr Q=\Delta\partial_r^2+{}$(terms with at most one radial derivative), the product rule gives, for $f\in\{U,V\}$ and $i\geq2$,
\begin{equation} \label{eq:supp-triangular-division}
    \partial_r^i\partial_x^jf
    =\frac{1}{\Delta}
    \Big[\partial_r^{i-2}\partial_x^j(\mathscr Qf)
    -\Xi_{ij}[f]\Big]\,,\qquad
    \Xi_{ij}[f]
    \equiv\partial_r^{i-2}\partial_x^j(\mathscr Qf)
    -\Delta\,\partial_r^i\partial_x^jf\,,
\end{equation}
where $\Xi_{ij}[f]$ contains only derivatives of $f$ with fewer than $i$ radial derivatives. Applying this replacement from $i=4$ down to $i=2$ removes every derivative of $U$ and $V$ with two or more radial derivatives in favor of derivatives of $\mathscr QU=\operatorname{Re}\mathscr QF$ and $\mathscr QV=\operatorname{Im}\mathscr QF$ (recall that $\mathscr Q$ is real). After this replacement, the coefficients of all remaining derivatives of $U$ and $V$ vanish identically in all ten components, for arbitrary $U$ and $V$. So we have
\begin{equation} \label{eq:supp-einstein-right-factorization}
    \delta G_{\mu\nu}[h^{\rm pot}]
    =\mathcal C_{\mu\nu}[\mathscr Q F]\,,\qquad
    \mathcal C_{\mu\nu}[Z]
    =\sum_{i+j\leq2}\left[c^{\rm R}_{\mu\nu ij}
    \,\partial_r^i\partial_x^j\operatorname{Re}Z
    +c^{\rm I}_{\mu\nu ij}
    \,\partial_r^i\partial_x^j\operatorname{Im}Z\right]\,,
\end{equation}
where the rational coefficients $c^{\rm R}_{\mu\nu ij}$ and $c^{\rm I}_{\mu\nu ij}$ do not depend on $F$. Thus, the Einstein tensor of the potential part is a fixed second-order operator acting on $\mathscr QF$. In particular, $h^{\rm pot}$ solves the linearized vacuum equations whenever $\mathscr QF=0$, as in Wald's adjoint construction~\cite{Wald:1978vm}.

The rational part is evaluated directly. Inserting $h^{\rm rat}$ into Eq.~\eqref{eq:supp-coordinate-einstein} gives the ten identities $\delta G_{\mu\nu}[h^{\rm rat}]+\mathcal C_{\mu\nu}[W]=\mathcal S_{\mu\nu}$. For the truncated metric $h_{\lmax}^{(1)}=h^{\rm pot}|_{F=F_{\lmax}}+h^{\rm rat}$, $\mathscr QF_{\lmax}=W_{\lmax}$, so
\begin{equation}
       \delta G_{\mu\nu}[h_{\lmax}^{(1)}]-\mathcal S_{\mu\nu}
       =\mathcal C_{\mu\nu}[W_{\lmax}]-\mathcal C_{\mu\nu}[W]=\mathcal C_{\mu\nu}[W_{\lmax}-W]\,.
 \label{eq:supp-finite-einstein-defect}
\end{equation}
This identity holds for all $M$, $a$, $r$, and $x$. The Einstein residual of the truncated metric is therefore set entirely by the source multipoles omitted by the truncation, $W-W_{\lmax}$, whose coefficients $\mathcal F_\ell$ contain only rational functions and $L$.

%%%%%%%%%%%%%%%%%%%%%%%%%%%%%%%%%%%%%%%%%%%%%%%%%%%%%%%%%%%%%%
%%%%%%%%%%%%%%%%%%%%%%%%%%%%%%%%%%%%%%%%%%%%%%%%%%%%%%%%%%%%%%
\subsection{Angular convergence rate} 
\label{sec: angular convergence}

We now show that the residual decay exponentially in $\ell$. For $r\geq r_+$, both angular sources, $\mathcal T_2$ and $Q_{\rm full}$, are $(1-x^2)$ times a rational function of $x$ whose only singularity is a pole at $\Gm=0$, i.e., $x_0=ir/a$ (of order 11 and 7, respectively). Their coefficients $T_\ell$ and $\mathcal F_\ell$ in the basis $P_\ell^2=(1-x^2)P_\ell''$ follow from two integrations by parts,
\begin{equation} \label{smtc:coefficient-ibp}
    \nu_\ell\int_{-1}^{1}P_\ell^2\,g\,dx
    =\nu_\ell\int_{-1}^{1}P_\ell\,\big[(1-x^2)g\big]''\,dx\,,
    \qquad g\in\{\mathcal T_2,Q_{\rm full}\}\,,
\end{equation}
where the boundary terms vanish because $(1-x^2)g$ contains $(1-x^2)^2$. Thus, $T_\ell$ and $\mathcal F_\ell$ are, up to $\nu_\ell=\mathcal O(\ell^{-3})$, ordinary Legendre coefficients of rational functions whose only singularity is a pole at $x_0$. The pole of the rational function determines the large $\ell$ behavior of the Legendre coefficients. For a simple pole, Heine's formula
\begin{equation} \label{smtc:heine}
    \frac{1}{z-x}
    =\sum_{\ell\geq0}(2\ell+1)P_\ell(x)\,Q_\ell(z)\,,\qquad
    Q_\ell(z)
    \simeq\sqrt{\frac{\pi}{2\ell}}\,(z^2-1)^{-1/4}
    \big[z+\sqrt{z^2-1}\big]^{-\ell-1/2}\quad(\ell\to\infty)\,,
\end{equation}
gives the coefficients exactly. Poles of higher order are $z$ derivatives of it and only add powers of $\ell$. At $z=x_0$, the bracket has modulus $\varrho=|x_0+\sqrt{x_0^2-1}|=(r+\sqrt{r^2+a^2})/|a|$, so both $\mathcal F_\ell$ and $T_\ell$ decay as $\varrho^{-\ell}$ up to powers of $\ell$. The decay is slowest at the horizon, where $r_+^2+a^2=2Mr_+$ gives $\varrho_+=(r_++\sqrt{2Mr_+})/|a|$. The decay at the horizon decreases monotonically in $|a|$ to $1+\sqrt2$ at $|a|=M$. The Einstein residual~\eqref{eq:supp-finite-einstein-defect} involves only $W-W_{\lmax}$, is local in $r$, and therefore decays as $\varrho(r)^{-\lmax}$.

%TC:endignore

% END SUPPLEMENTAL MATERIAL
\end{document}